\documentclass[letterpaper,twocolumn,10pt]{article}
\usepackage{usenix-2020-09}

\usepackage{authblk}
\usepackage{tikz}
\usepackage{amsmath}
\usepackage{amssymb}
\usepackage{pifont}
\usepackage{filecontents}

\usepackage{caption}
\usepackage{graphicx}
\usepackage{diagbox}
\usepackage{float} 
\usepackage{subcaption}
\usepackage{threeparttable}
\usepackage{multirow}
\usepackage[linesnumbered,lined,boxed,commentsnumbered]{algorithm2e}

\newtheorem{theorem}{Theorem}
\newtheorem{lemma}{Lemma}

\newtheorem{definition}{Definition}

\newenvironment{thmprime2}[1]
  {\renewcommand{\thetheorem}{\ref{#1}}%
   \addtocounter{theorem}{-1}%
   \begin{theorem}}
  {\end{theorem}}

\makeatletter

\newcommand{\Rmnum}[1]{\expandafter\@slowromancap\romannumeral #1@}
\makeatother

\usepackage [
    n, 
    advantage,
    operators,
    sets,
    adversary,
    landau,
    probability,
    notions,
    logic,
    ff, 
    mm,
    primitives,
    events,
    complexity,
    oracles,
    asymptotics,
    keys
]{cryptocode}
\usepackage{enumitem}

\begin{document}

\date{}

\title{\Large \bf Arbitrary-Threshold Fully Homomorphic Encryption with Lower Complexity}

\author[$\dag$]{Yijia Chang}
\author[$\,\,\,\,\,$]{Songze Li\thanks{Corresponding author: Songze Li.}}
\affil[$\dag$]{The Hong Kong University of Science and Technology (Guangzhou)$\qquad$\textsuperscript{*}{Southeast University}}
\affil[ ]{\textit{ychang847@connect.hkust-gz.edu.cn, songzeli8824@outlook.com}}

\maketitle

\begin{abstract}
    Threshold fully homomorphic encryption (ThFHE) enables multiple parties to compute functions over their sensitive data without leaking data privacy. Most of existing ThFHE schemes are restricted to full threshold and require the participation of \textit{all} parties to output computing results. Compared with these full-threshold schemes, arbitrary threshold (ATh)-FHE schemes are robust to non-participants and can be a promising solution to many real-world applications. However, existing AThFHE schemes are either inefficient to be applied with a large number of parties $N$ and a large data size $K$, or insufficient to tolerate all types of non-participants. In this paper, we propose an AThFHE scheme to handle all types of non-participants with lower complexity over existing schemes. At the core of our scheme is the reduction from AThFHE construction to the design of a new primitive called \textit{approximate secret sharing} (ApproxSS). Particularly, we formulate ApproxSS and prove the correctness and security of AThFHE on top of arbitrary-threshold (ATh)-ApproxSS's properties. Such a reduction reveals that existing AThFHE schemes implicitly design ATh-ApproxSS following a similar idea called ``noisy share''. Nonetheless, their ATh-ApproxSS design has high complexity and become the performance bottleneck. By developing ATASSES, an ATh-ApproxSS scheme based on a novel ``encrypted share'' idea, we reduce the computation (resp. communication) complexity from $\mathcal{O}(N^2K)$ to $\mathcal{O}(N^2+K)$ (resp. from $\mathcal{O}(NK)$ to $\mathcal{O}(N+K)$). We not only theoretically prove the (approximate) correctness and security of ATASSES, but also empirically evaluate its efficiency against existing baselines.
    Particularly, when applying to a system with one thousand parties, ATASSES achieves a speedup of $3.83\times$ -- $15.4\times$ over baselines.

\end{abstract}

\section{Introduction}

Threshold fully homomorphic encryption (ThFHE) allows arbitrary computations over encrypted data from \textit{multiple} parties, without decrypting them~\cite{mpcviaFHE,mpcviaFHE0,systemImplementation,Threshold,efficient,interactiveTFHE}. Different from traditional single-party FHE, threshold FHE supports to distribute the data-controlling authority to multiple parties, i.e., the decryption can not succeed unless enough parties participate.
This provides more flexibility and induces a family of privacy-preserving computing schemes in the secure multiparty computation (MPC) setting, where multiple parties wish to evaluate a function over their joint inputs while ensuring the privacy of inputs~\cite{efficient}. Compared with classical MPC methods, these ThFHE-based solutions are distinguished by their low communication complexity~\cite{mpcviaFHE} and compatibility with cloud-assisted setting such as multiple-client-single-server architecture~\cite{systemImplementation}. To date, ThFHE has been implemented by open-source libraries~\cite{lattigo,palisade} and shown various promising real-world applications, including privacy-preserving machine learning~\cite{ApplicationFL, ApplicationFL2, ApplicationFL3, ApplicationFL4,interactiveTFHE} and medical analytics~\cite{ApplicationMedical,ApplicationMedical1,ApplicationMedical2,ApplicationMedical3, realworldApp}.

Along with the flexible controlling authority comes a new factor that should be taken into consideration: that is, how to set the threshold of participants required for decryption. Let $N$ denote the number of parties and $T$ denote such a threshold ($T\leq N$). Most of existing ThFHE schemes~\cite{mpcviaFHE,mpcviaFHE0,systemImplementation} are designed in the \textit{full-threshold} case with $T=N$, i.e., the decryption can succeed only if \textit{all} parties participate. Nonetheless, achieving full-threshold is often challenging in practice due to the following reasons.
\begin{itemize}
    \item[$\bullet$] \emph{Uninterested Parties.} The decryption process would incur high computation and communication costs of participants. Faced with these costs, a party may lose interest in the decryption result and quit the decryption process, because the result is less valuable than those costs or this party needs to handle tasks with higher priority.
    \item[$\bullet$] \emph{Dropout Devices.} In practice, parties are usually some devices (e.g., smart phones and sensors) that are interconnected as a network. These devices may drop out of network due to unexpected random factors (e.g., device powering-off and loss of network connectivity) and are unable to participate in the decryption.
    \item[$\bullet$] \emph{Denial-of-Service Attacks.} The aim of denial-of-service (DoS) attack is to make some ``service'' unavailable to its intended users. For ThFHE, a common goal of DoS attack is to thwart parties from obtaining decryption results. To launch DoS attack against full-threshold FHE schemes, an attacker needs to invade only one party and prevent it from participating in the decryption process.
\end{itemize}
These reasons result in various types of non-participating parties, which may fail the decryption process and degrade the practicality of full-threshold FHE schemes~\cite{Threshold,efficient}.
To improve the practicality of full-threshold FHE, a natural idea is to design \textit{arbitrary}-threshold FHE (AThFHE) that allows elastic choices of threshold $T$ from $1$ to $N-1$. In this way, AThFHE can still guarantee the output delivery of decryption protocol even when $N-T$ parties do not participate.

To this end, recent works have proposed several AThFHE schemes~\cite{Threshold,efficient,interactiveTFHE,nuoyafangzhou,realworldApp} based on linear secret sharing (LSS).
Notably, all of them follow the same idea called ``noisy share'' when devising the decryption protocols. Specifically, to prevent the secret key from leakage during the decryption process, each party needs to compute a decryption share and add a small noise to it.
However, to guarantee that the aggregation of these noisy decryption shares leads to correct decryption results, the decryption requires well-crafted design that usually incurs high complexity.
For example, Boneh et al. use LSS with small recovery coefficients~\cite{Threshold,realworldApp} to devise a AThFHE. Nonetheless, compared with single-party FHE, the decryption key share has the size of order $\mathcal{O}(N^{4.2})$ on average.
They also construct another AThFHE based on Shamir LSS, which has the optimal share-size.
Still, this scheme has a huge ciphertext space with modulus $\mathcal{O}(N\cdot(N!)^3)$~\cite{Threshold}, incurring high complexity and difficulty for parameter instantiation.
To improve the efficiency, some works propose decryption protocols with two rounds~\cite{interactiveTFHE,nuoyafangzhou}. The computation (resp. communication) complexity of decryption protocol is optimized to $\mathcal{O}(N^2\cdot K)$ (resp. $\mathcal{O}(N\cdot K)$), where $K$ is the size of decrypted message. Still, such a complexity is relatively high, especially when the number of parties and the length of message are both large. 
Mouchet et al. propose an idea to further reduce the complexity in~\cite{efficient} and extend this idea in Helium framework~\cite{Helium}. Nonetheless, this idea assumes that the set of participants is known prior to decryption, which may not hold when the non-participants are caused by random dropout devices or denial-of-service attacks.


In a word, following the ``noisy share'' idea, existing AThFHE schemes are either inefficient to be applied on large systems and large data size, or insufficient to handle all types of non-participants. These deficiencies severely restrict the application of AThFHE in the real world. Based on the above investigation, we are motivated to ask the questions:
\textit{Is this ``noisy share'' idea the only and the optimal way to design AThFHE? If not, how can we further reduce its complexity?}

\begin{figure}[tb]
	\centering
	\includegraphics[width=0.95\linewidth]{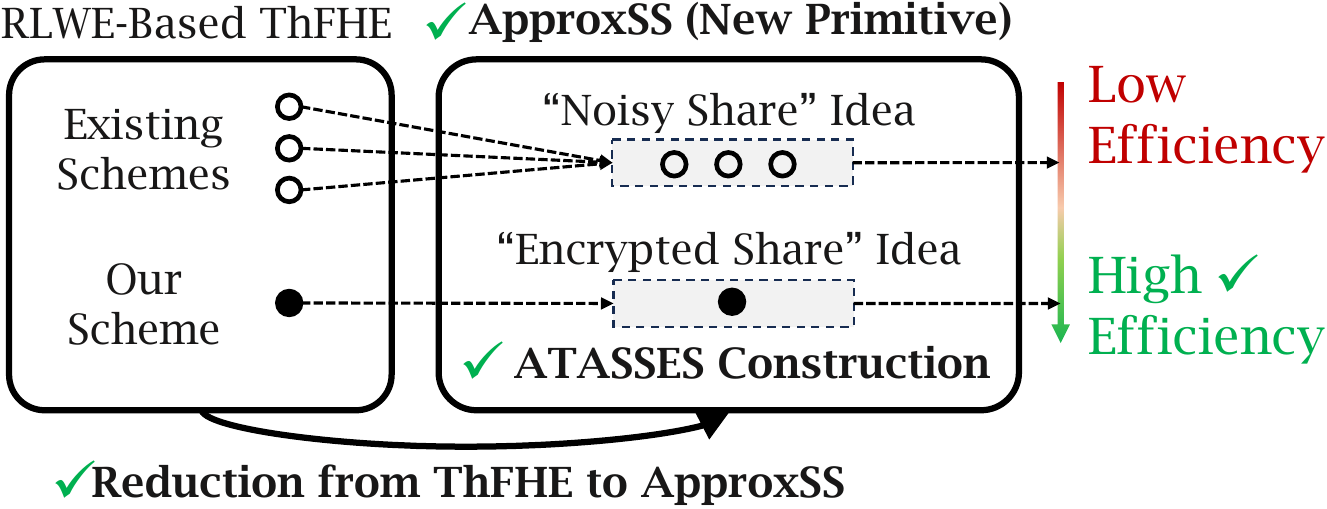}
	\caption{An overview of our main results.}
	\label{fig:framework}

\end{figure}

In this work, we answer these questions by reducing the AThFHE design to the construction of \textit{approximate} secret sharing (ApproxSS), a novel cryptographic primitive formulated in this paper.
The ``noisy share'' idea followed by existing AThFHE schemes can be regarded as a specific type of construction for ApproxSS. 
We propose a novel idea called ``encrypted share'' to construct ApproxSS with computation complexity of $\mathcal{O}(N^2+NK)$ and communication complexity of $\mathcal{O}(N+K)$, which achieves order-wise improvements over existing AThFHE schemes. Below we summarize our main results and key contributions, and illustrate them in Figure \ref{fig:framework}.

\begin{itemize}[leftmargin=*]
    \item \textbf{Formulation of Approximate Secret Sharing}. Beyond vanilla secret sharing, a $T$-out-of-$N$ ApproxSS has an additional operation called approximate recovery, through which any $T$ shares can recover an approximation of the secret (a.k.a., approximate correctness), while the adversary corrupting up to $T-1$ parties can NOT learn the exact value of secret through approximate recovery (a.k.a., approximate security). Although existing works implicitly construct several ApproxSS schemes, to the best of our knowledge, this is the first work to formulate its definition and security. We believe this primitive can be of independent interest for other applications, such as the intersection of secure multi-party computation and differential privacy.
    \item \textbf{Reduction from RLWE-Based ThFHE to ApproxSS.} We propose a generic construction of ring-learning-with-errors (RLWE)-based ThFHE on top of ApproxSS and prove its security and correctness based on ApproxSS's properties. Particularly, on input decryption shares from any $T$ parties, the approximate recovery in ApproxSS can output an approximate message. This approximate message can not only yield the correct plaintext by the feature of RLWE-based HE, but also protect the secret keys with the well-known smudging technique~\cite{mpcviaFHE}. Notably, existing AThFHE schemes are special cases of this generic construction. Nonetheless, their ApproxSS design is less efficient and becomes the performance bottleneck of ThFHE.

    \item \textbf{An Efficient Arbitrary-Threshold ApproxSS.} In light of the above generic construction, we propose ATASSES, an Arbitrary-Threshold ApproxSS based on the idea of ``Encrypted Share''. This idea protects decryption shares using crafted encryption methods rather than adding noise, which helps to boost Shamir SS to ApproxSS in an efficient manner. We prove the approximate correctness and security of ATASSES, and show that ATASSES achieves the computation complexity of $\mathcal{O}(N^2+NK)$ and the communication complexity of $\mathcal{O}(N+K)$. 
    \item \textbf{Performance Evaluation through Experiments.} We implement ATASSES and existing baselines on top of Lattigo library~\cite{lattigo}. Experimental results validate the superior performance of ATASSES. Particularly, when applying to a system with one thousand parties, ATASSES achieves a speedup of $3.83\times$ -- $15.4\times$ than state-of-the-art ApproxSS. Such a speedup can be even more significant with more parties or larger data size.
\end{itemize}
The rest of this paper is organized as follows. We describe our system and threat models in Section \ref{sec:thmphe-0}. Then we provide our technical intuition in Section \ref{sec:intuition} and necessary preliminaries in Section \ref{sec:pre}. Based on these knowledge, Section \ref{sec:thmphe} formulates ApproxSS and shows the generic construction of ThFHE on top of ApproxSS. Next, Section \ref{sec:ass} proposes ATASSES and Section \ref{sec:experiment} empirically evaluates its efficiency.
At last, Section \ref{sec:related} reviews related work and Section \ref{sec:conclusion} concludes this paper.

\section{System and Threat Model}
\label{sec:thmphe-0}

\begin{figure}[htb]
	\centering
	\includegraphics[width=0.9\linewidth]{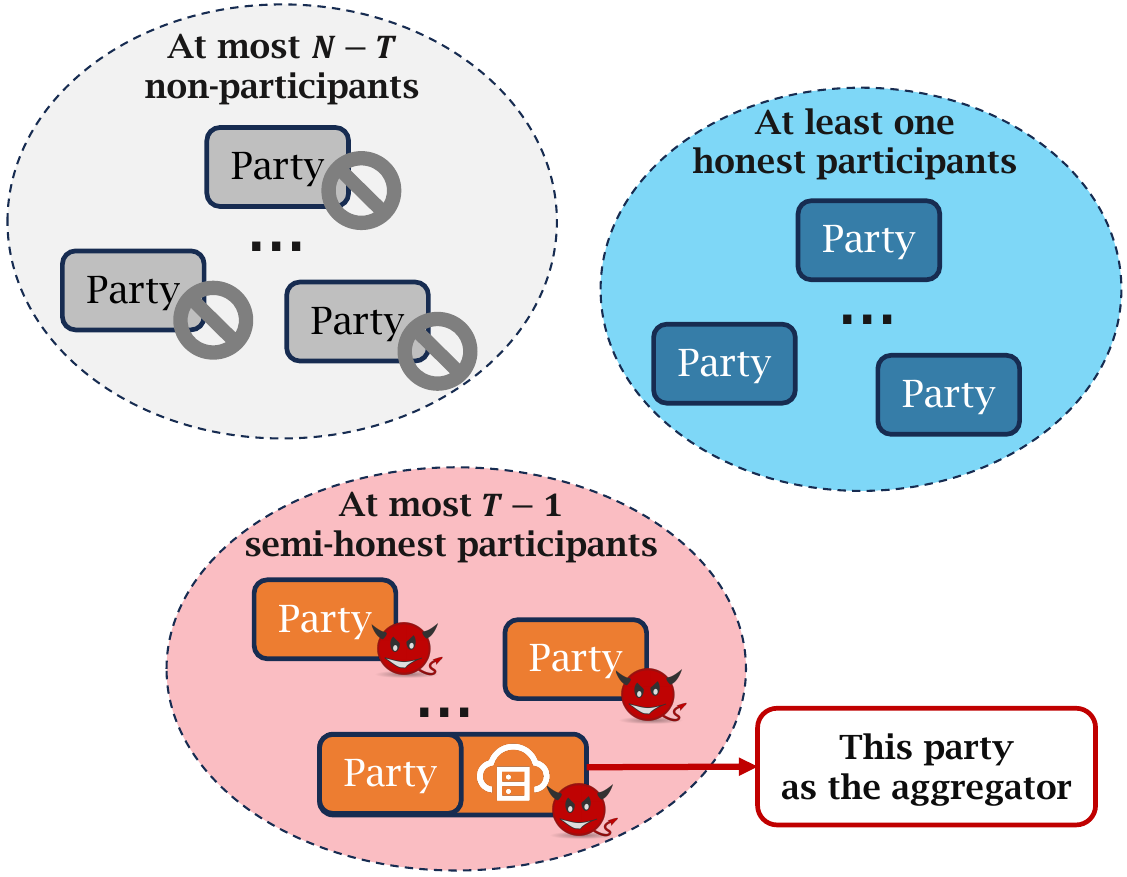}
	\caption{System model with three types of parties.}
	\label{fig:system}
\end{figure}

\noindent\textbf{System Model.} As illustrated in Figure \ref{fig:system}, we consider a group of $N$ parties who wish to employ threshold FHE to jointly compute a function $f(\cdot)$ over their sensitive data. 
To facilitate the execution of threshold FHE, we assume an \textit{aggregator} who combines cryptographic transcripts from parties and homomorphically evaluates the function $f(\cdot)$ over ciphertexts. This role can be played by any of available parties or an external server, such as the cloud. The parties (and the aggregator) are interconnected via authenticated channels.
The aggregator would output one or more FHE ciphertexts, so that decrypting them leads to the desired output of function $f(\cdot)$.
Throughout this paper, we use $K$ to denote the length of function $f(\cdot)$'s output.
We also assume the setting of \textit{common reference string} (CRS), where parties can generate the same random value $a$ from some distribution. We use $a\sample\mathsf{CRS}$ to denote this case. 

In practice, some parties may fail to participate in the decryption protocol due to losing interests, dropout devices, and/or denial-of-service attacks. We call these parties ``non-participants'' and other parties ``participants''. Nonetheless, we assume that there are at least $T$ participants at the same time. More formally, when the decryption protocol asks parties to send some message to other parties, there are at least $T$ parties who can upload the message within the required time. In addition, we require that a party can NOT know the set of participants until it receives the demanded messages from participants. The reason is that the non-participants may be unable to inform other parties due to random factors or even intentional attacks. In these cases, no one can assure which parties will succeed to upload the demanded messages. We remark that this is one key difference between our setting and existing work~\cite{efficient}, which makes our work more practical.


\noindent\textbf{Threat Model.} We consider a static semi-honest adversary who can corrupt a fixed set of up to $T-1$ parties, \textit{i.e.}, these $T-1$ parties faithfully run the algorithms and protocols in ThFHE schemes, but the adversary can see their internal states to infer private messages. Such a setting implies that there exists at least one honest participant. In addition, we do not assume a trustworthy aggregator, i.e., the aggregator is corrupted by the adversary. 

We note that our threat model, combined with the non-participants setting in the system model, is in fact a mixed adversary setting. Particularly, those non-participants can be regarded as being corrupted by a partly-malicious adversary, who can only abort at most $N-T$ parties simultaneously. We also note that recent work~\cite{PELTA} makes the first move to shield the ThFHE schemes against malicious adversaries. Despite of its great success, they assume that the parties do not refuse to participate in the protocol execution. Since our work considers the non-participant problem, we believe our work can help to remove this assumption and boost ThFHE schemes with fully-malicious security.
We provide a discussion on how to extend our scheme to fully-malicious security in Appendix \ref{appendix:fullysecurity}.

\section{Technical Intuition}
\label{sec:intuition}

This section introduces the intuition behind our techniques on a high level.
For brevity's sake, the notation in this section are highly intuitive but somewhat informal.

\noindent\textbf{Problem Description.}
At the core of $T$-out-of-$N$ ThFHE construction is to design a decryption protocol that allows any $T$ participants to decrypt without disclosing the secret key.
Suppose that the secret key is $\mathsf{sk}$ and the ciphertext is $\mathsf{ct}$, the decryption algorithm in existing FHE schemes is usually to compute a linear function $b=\mathsf{ct}\cdot\mathsf{sk}$. In current lattice-based FHE schemes, the $\mathsf{ct}$ contains an error $e_{ct}$. 
The plaintext can be successfully decoded from $b$ by rounding and/or modular reduction if the value of $e_{ct}$ is less than some bound $B$.

Since the decryption is to compute a linear operation, existing ThFHE construction usually applies \textit{linear secret sharing} technique to $\mathsf{sk}$. This will generate $N$ key shares $\mathsf{sk}_1,\ldots,\mathsf{sk}_N$, one for each party. To decrypt $\mathsf{ct}$, each party $i$ computes a decryption share $b_i=\mathsf{ct}\cdot\mathsf{sk}_i$. By the linearity of secret sharing, decryption shares $b_i$s are the shares of $b$, and the aggregator can recover $b$ from any $T$ decryption shares. 
For example, the full-threshold ThFHE adopts \textit{additive} secret sharing so that $\mathsf{sk}=\sum_{i=1}^N \mathsf{sk}_i$. After receiving $N$ decryption shares ($T=N$), the aggregator can compute $\sum_{i=1}^N b_i=\mathsf{ct}\cdot\sum_{i=1}^N \mathsf{sk}_i=\mathsf{ct}\cdot \mathsf{sk}=b$.
However, this construction is insecure, as $b$ leaks information of $\mathsf{sk}$ to the aggregator. To solve this issue, a common solution is noise smudging technique~\cite{mpcviaFHE}. Specifically, the decryption protocol can recover $b^\prime=b+n_{sm}$ for some random smudging noise $n_{sm}$ rather than $b$. As long as the value of $n_{sm}+e_{ct}$ is bounded by $B$, the decryption can still succeed.

To do so, existing ThFHE constructions ask each party $i$ to sample a local noise $n_{sm,i}$ and compute a noisy share $b_i^\prime=b_i+n_{sm,i}$. We call this idea ``noisy share''. 
By the linearity of secret sharing, noisy shares $b_i^\prime=b_i+n_{sm,i}$ can be regarded as the shares of $b+n_{sm}$ for some $n_{sm}$ that is determined by $n_{sm,i}$s. In existing ThFHE constructions, the aggregator recovers $b+n_{sm}$ as $b^\prime$ from any $T$ noisy shares.
For example, in the full-threshold case, the recovery algorithm is simply summing all shares up. Then the aggregator will compute $b^\prime=\sum_{i=1}^N b_i^\prime=b+\sum_{i=1}^N n_{sm,i}=b+n_{sm}$, where $n_{sm}=\sum_{i=1}^N n_{sm,i}$.

Still, this ``noisy share'' idea faces a challenge. Recall that the value of $n_{sm}+e_{ct}$ should be bounded by $B$ for successful decryption. Hence, this idea needs to guarantee that the value of $n_{sm}$ is bounded by $B_{sm}$ with $B_{sm}+\norm{e_{ct}}<B$. For full-threshold FHE, this challenge can be easily tackled by setting the value of each local noise $n_{sm,i}$ bounded by $B_{sm}/N$. Under this setting, we have $\norm{n_{sm}}<\sum_{i=1}^N \norm{n_{sm,i}}<B_{sm}$. Nonetheless, this challenge becomes particularly hard for arbitrary-threshold FHE. Note that $n_{sm,i}$s are locally sampled by each party $i$ and have no natural correlation. Without crafted design on $n_{sm,i}$s, the corresponding $n_{sm}$ may have random values without bound. For example, when using the well-known Shamir secret sharing, we have $n_{sm}=\sum_{i\in\mathcal{T}} L^{(\mathcal{T})}_i n_{sm,i}$, where $\mathcal{T}$ is a set of at least $T$ parties and $L^{(\mathcal{T})}_i$s are Lagrange coefficients associated with set $\mathcal{T}$. Since Lagrange coefficients have arbitrarily-large value, they may blow the noise $n_{sm}$ up and fail the decryption.

Following the ``noisy share'' idea, existing ThFHE schemes try to solve this ``blown-up noise'' problem by restricting the value of $n_{sm,i}$ in various ways. Unfortunately, these solutions either suffer from high complexity or rely on an additional assumption. 
Particularly, without assuming that the knowledge of participant set $\mathcal{T}$ is available before decryption, the state-of-the-art solution~\cite{interactiveTFHE,nuoyafangzhou} has the computation complexity of $\mathcal{O}(N^2K)$ and communication complexity of $\mathcal{O}(NK)$.
Readers can find more detailed discussion in Section \ref{sec:ass-existing}.

\begin{figure}[htb]
	\centering
	\includegraphics[width=0.8\linewidth]{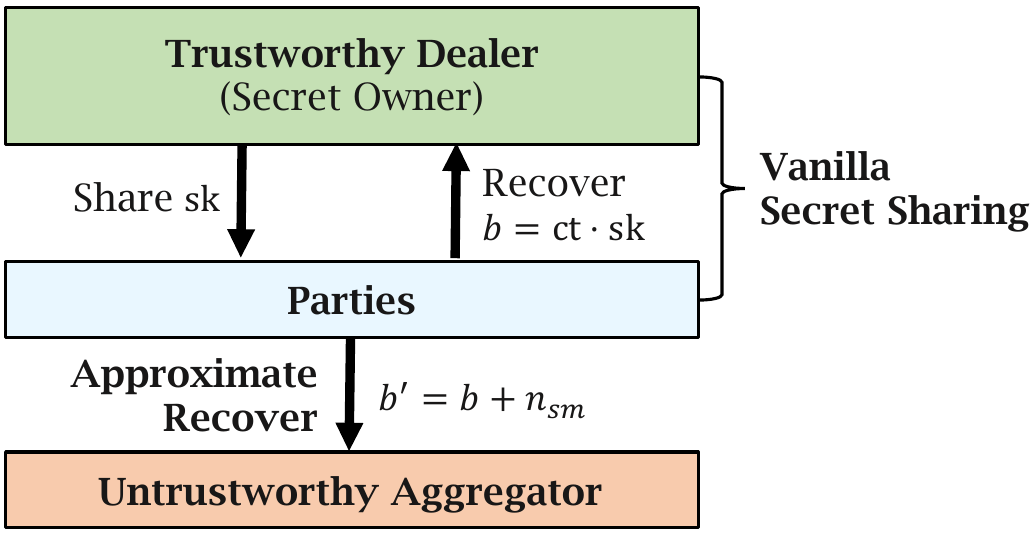}
	\caption{An illustration of approximate secret sharing.}
	\label{fig:approxSS}
\end{figure}

\noindent\textbf{Our Solution.} Our first observation is that ``noisy share'' is not the only way to apply the noise smudging technique for ThFHE construction. Basically, the noise smudging technique inspires us that recovering the exact value of $b$ is neither secure nor necessary. Instead, the aggregator only needs to recover an approximate value $b^\prime=b+n_{sm}$ so that 1) $n_{sm}$ has a bounded value and 2) $n_{sm}$ is random that cannot be determined by less than $T$ parties. We realize that vanilla secret sharing is insufficient to meet these two requirements and thus call for a novel cryptographgic primitive. As shown in Figure \ref{fig:approxSS}, vanilla SS only enables the trustworthy dealer to exactly recover the secret (i.e., $b$), while now an untrusted aggregator needs to recover the secret in an approximate manner (i.e., $b^\prime$).

Hence, beyond the original functions of vanilla secret sharing, the new primitive should support a novel function called ``approximate recovery''. This function should enable the aggregator to learn the value of $b^\prime$ from any $T$ shares of $b$, such that 1) $b^\prime$ is ``close'' to $b$ in the sense that $\norm{b^\prime-b}$ is bounded and 2) the adversary corrupting less than $T$ parties cannot determine the value of $b^\prime-b$.
We term this primitive \textit{approximate secret sharing} (ApproxSS) and formulate it in Section \ref{sec:ass-def}.
We further show that a $T$-out-of-$N$ ThFHE scheme can be constructed based on a $T$-out-of-$N$ ApproxSS scheme in Section \ref{sec:thmphe-2}. Notably, the noisy share idea is an implicit construction of this primitive. However, their ApproxSS constructions has become the performance bottleneck of ThFHE scheme due to the high complexity. This inspires us to study more efficient ApproxSS construction. 

Our second observation is another way to boost Shamir secret sharing to ApproxSS. 
This idea relies on an encryption scheme that can enable the following three steps.

\noindent\textit{Step 1}: Each party $i$ locally encrypts $b_i$ and $n_{sm,i}$, respectively, and sends the ciphertexts to the aggregator.

\noindent\textit{Step 2}: After receiving the ciphertexts of $b_i$s and $n_{sm,i}$s from at least $T$ parties in $\mathcal{T}$, the aggregator will learn the set $\mathcal{T}$ and further compute the corresponding $L_i^{(\mathcal{T})}$s. The encryption scheme should be somewhat linearly-homomorphic, so that the aggregator can obtain a new ciphertext of $b^\prime=\sum_{i\in\mathcal{T}} (L_i^{(\mathcal{T})} b_i+n_{sm,i})$ from the ciphertexts of $b_i$s and $n_{sm,i}$s.

\noindent\textit{Step 3}: Any $T$ parties collaboratively decrypt the ciphertext of $b^\prime$, without leaking individual $b_i$ or $n_{sm,i}$. Notably, these $T$ parties can be totally different from the set $\mathcal{T}$ in the first step.

We call this idea ``encrypted share'' and provide a specific construction based on BFV secret-key encryption in Section \ref{sec:ass}. We argue that this idea has at least the following three advantages, especially against the noisy share idea.

\begin{itemize}[leftmargin=*]
    \item Our idea avoids the ``blown-up noise''. By encrypting $b_i$ and $n_{sm,i}$, respectively, the aggregator can compute their linear combination with different coefficients, e.g., $L_i^{(\mathcal{T})}\cdot b_i+n_{sm,i}$. For comparison, in the noisy share idea, the $b_i+n_{sm,i}$ is uploaded to the aggregator as a whole. If the aggregator multiplies it with $L_i^{(\mathcal{T})}$ to recover $b$, the $n_{sm,i}$ is also multiplied by $L_i^{(\mathcal{T})}$, resulting in the blown-up noise. In addition, we note that in our idea, the Lagrange coefficients will blow up the errors in BFV ciphertexts. If the error exceeds some bound, the decryption also fails. Nonetheless, as we will show in Section \ref{sec:ass-our}, this problem can be solved by selecting a slightly larger ciphertext space with modulus $\mathcal{O}(N)$.
    \item Although our idea also requires to share and recover (encryption and decryption) keys with $\mathcal{O}(N^2)$ computation complexity and $\mathcal{O}(N)$ communication complexity, these expensive operations only involve the keys whose size is a \textit{constant} independent from the length of ciphertexts $K$. As a result, the total computation (resp. communication) complexity is $\mathcal{O}(N^2+NK)$ (resp. $\mathcal{O}(N+K)$). For comparison, the state-of-the-art solution~\cite{interactiveTFHE,nuoyafangzhou} following the noisy share idea requires to share and recover noises whose length is the same as ciphertexts and thus its computation (resp. communication) complexity is $\mathcal{O}(N^2K)$ (resp. $\mathcal{O}(NK)$).
    \item Our idea does not require to learn the set of participants in advance. In Step 2, only \textit{after} the aggregator has received the ciphertexts from the parties in $\mathcal{T}$, the aggregator needs to know the set $\mathcal{T}$ and $L_i^{(\mathcal{T})}$s. 
    Moreover, the participants in Step 1 are allowed not to participate Step 3. These settings match our system model on (non)-participants. Despite an existing work~\cite{efficient,Helium} based on the noisy share idea has lower complexity, it relies on an extra assumption than our idea, i.e., the prior knowledge of participant set is available.
\end{itemize}

By the theoretical analysis in Section \ref{sec:ass-analysis} and empirical evaluation in Section \ref{sec:experiment}, we verify the superior performance of our BFV-based construction. Still, our construction may not be the optimal one following the encrypted share idea. We expect future works to continuously improve its efficiency by using more ingenious encryption scheme. 

\section{Preliminaries}
\label{sec:pre}
\subsection{Notation}

For a number $M$ of two-power, this paper considers polynomial rings $\mathcal{R}_{M}=\mathbb{Z}[x]/(x^M+1)$ and $\mathcal{R}_{M,Q}=\mathcal{R}_{M}/Q\mathcal{R}_{M}$, and adopts bold letters to denote polynomials (e.g., $\mathbf{s}$). For a polynomial $\mathbf{s}$, its coefficient is denoted by $s_i$ and its infinity norm is defined as $\norm{\mathbf{s}}=\max_i|s_i|$.
The sets are denoted by calligraphic letters (e.g., $\mathcal{S}$) except two exceptions. One is $[N]=\{1,\cdots,N\}$ for integer $N$. The other is $\mathsf{negl}$ that denotes the set of negligible functions (with respect to some security parameter). In addition, $a\sample \chi$ and $a\sample \mathcal{M}$ denote that $a$ is sampled from a distribution $\chi$ or the uniform distribution over the set $\mathcal{M}$, respectively. We also use $\mathsf{Uniform}(\mathcal{M})$ to denote the uniform distribution over the set $\mathcal{M}$. 


\subsection{Vanilla Secret Sharing}
\label{sec:pre-SS}
Secret sharing (SS) enables a dealer to distribute its secret message $m$ to other parties in a secure manner~\cite{shamir1979share}. In this work, we consider $T$-out-of-$N$ secret sharing. Its goal is to enable any set of at least $T$ shares to recover the message $m$ (i.e., correctness), while prevent the set of less than $T$ shares from leaking anything about $m$ (i.e., security). The definition of secret sharing is provided as follows.

\begin{definition}[Vanilla Secret Sharing]
Given a message space $\mathcal{M}$, a $T$-out-of-$N$ secret sharing scheme is a pair of PPT algorithms $(\mathtt{Share},\mathtt{Rec})$ defined as follows:


\noindent\textbf{Share}: $\{s_i\}_{i\in[N]}$$\gets$$\mathtt{Share}(m)$. On input a message $m\in\mathcal{M}$, this algorithm outputs $N$ shares $\{s_i\}_{i\in[N]}$;

\noindent\textbf{Recovery}: $m\gets\mathtt{Rec}(\{s_i\}_{i\in\mathcal{T}})$. On input shares $\{s_i\}_{i\in\mathcal{T}}$ from a set $\mathcal{T}$, this algorithm outputs a message $m\in\mathcal{M}$.

\end{definition}

In addition, some SS schemes satisfy a special property called linearity~\cite{LISS}.
This property allows each share to contain multiple pieces and enables the message to be recovered as the linear combination of pieces from at least $T$ shares. More formally, the recovery of message $m$ is done as $m = \sum_{i\in\mathcal{T}}\sum_{l\in\mathcal{L}_i} w_{i,l}\cdot s_{i,l}$, where $\mathcal{T}$ is the set of at least $T$ shares, $\mathcal{L}_i$ is the set of pieces in $s_i$, $s_{i,l}$ is the $l$-th piece of $s_i$, and $w_{i,l}$ is called the recovery coefficients. Linearity enables a convenient method to compute the share of the linear combination of messages. Particularly, given that the $i$-th share of messages $\{m_1,\ldots,m_K\}$ is $\{s_{1,i},\ldots,s_{K,i}\}$, the $i$-th share of another message $m^\prime = \sum_{k\in[K]} b_k m_k$ can be computed as $s^\prime_i = \sum_{k\in[K]} b_k s_{k,i}$.

This work mainly uses Shamir SS~\cite{shamir1979share}, in which each share has only one piece. Suppose that each party $i$ owns a share $s_i$, then the message $m$ is recovered as $m=\sum_{i\in\mathcal{T}} L_i\cdot s_i$.
Here, $\mathcal{T}$ can be any set of at least $T$ parties and $L^{(\mathcal{T})}_i$s are called Lagrange coefficients that are computed as $L^{(\mathcal{T})}_i=\prod_{j\in\mathcal{T},j\neq i} [x_j/(x_j-x_i)]$, where $x_i$ is a number corresponding to party $i$.

\subsection{RLWE-Based Homomorphic Encryption}
\label{sec:pre-HE}
Homomorphic encryption (HE) is a special type of encryption that permits computations over ciphertexts without decryption. In this work, we focus on the HE schemes whose security is based on the hardness of ring-learning-with-errors (RLWE) problem, such as BGV~\cite{bgv}, BFV~\cite{bfv2012}, and CKKS~\cite{ckks,ckks2,ckks3}. These schemes can support additions and a bounded number of multiplications with high parallelized efficiency. By using the bootstrapping technique, these schemes can even evaluate an unbounded number of multiplications. Below we present BFV scheme. This scheme will be used as a concrete instance throughout this paper to introduce our scheme. Nonetheless, our work can be applied to other RLWE-based schemes, or even be naturally extended to other lattice-based HE schemes, such as Torus-FHE~\cite{torus} and FHEW~\cite{Ducas2015FHEWBH}.

In practice, the plaintext space and ciphertext space of BFV are polynomial rings $\mathcal{R}_{M,P}$ and $\mathcal{R}_{M,Q}$, respectively.
BFV involves an error distribution $\chi_{e,B}$, which is usually a discretized Gaussian distribution with norm bounded by $B$. Next, we describe the concrete algorithms of BFV scheme.

\noindent\textbf{Key Generation:} $\mathsf{sk},\mathsf{pk},\mathsf{evk}\gets \mathtt{KGen}()$. This algorithm outputs secret key $\mathsf{sk}$, public key $\mathsf{pk}$, and evaluation key $\mathsf{evk}$.

\noindent\textbf{Secret-Key Encryption:} $\mathsf{CT}\gets \mathtt{SKEnc}(\mathsf{sk},\mathbf{m},\mathbf{a})$. On input $\mathsf{sk}$, a message $\mathbf{m}\in\mathcal{R}_{M,P}$, and a random polynomial $\mathbf{a}\sample\mathcal{R}_{M,Q}$, this algorithm outputs a pair $\mathsf{CT}=(\mathsf{CT}[0],\mathsf{CT}[1])$ as the ciphertext. 

{
\sloppy
\noindent\textbf{Public-Key Encryption:} $\mathsf{CT}\gets \mathtt{PKEnc}(\mathsf{pk},\mathbf{m})$. On input $\mathsf{pk}$ and a message $m\in\mathcal{R}_{M,P}$, this algorithm outputs a pair $\mathsf{CT}=(\mathsf{CT}[0],$ $\mathsf{CT}[1])$ as the ciphertext.
}

{
\sloppy
\noindent\textbf{Homomorphic Evaluation:} $\mathsf{CT}^\prime\gets\mathtt{Eval}(\mathsf{evk},\{\mathsf{CT}_i\}_{i\in[N]}$,
$f(\cdot))$. On input $\mathsf{evk}$, $N$ ciphertexts $\{\mathsf{CT}_i\}_{i\in[N]}$, and an $N$-input function $f(\cdot)$, this algorithm outputs a new ciphertext $\mathsf{CT}^\prime$.
}

\noindent\textbf{Decryption:} $\mathbf{m}\gets\mathtt{Dec}(\mathsf{sk},\mathsf{CT})$. On input $\mathsf{sk}$ and ciphertext $\mathsf{CT}\in\mathcal{R}_{M,Q}$, this algorithm outputs a message $\mathbf{m}\in\mathcal{R}_{M,P}$.

    In the BFV scheme, the ciphertext $\mathsf{CT}=(\mathsf{CT}[0],\mathsf{CT}[1])$ of a message $\mathbf{m}$ under secret key $\mathsf{sk}$ has the structure of $\mathsf{CT}[0]+\mathsf{CT}[1]\cdot \mathsf{sk}=\Delta\cdot \mathbf{m}+\mathbf{e}_{\mathsf{CT}}$, where $\Delta=\lfloor Q/P\rfloor$ and $\mathbf{e}_{\mathsf{CT}}$ is the error in ciphertext. For example, the secret key encryption $\mathtt{SKEnc}(\mathsf{sk},\mathbf{m},\mathbf{a})$ is performed by sampling an error $\mathbf{e}\sample\chi_{e,B}$ and computing $\mathsf{CT}[0]=\mathbf{a}\cdot\mathsf{sk}+\Delta \cdot \mathbf{m}+\mathbf{e}$ and $\mathsf{CT}[1]=-\mathbf{a}$.

\noindent\textbf{Decryption and Its Requirement.} By the ciphertext structure, the decryption is done in two steps. The first step is to compute $\mathbf{b}=\mathsf{CT}[0]+\mathsf{CT}[1]\cdot \mathsf{sk}$, which is equal to $\Delta\cdot \mathbf{m}+\mathbf{e}_{\mathsf{CT}}$. The second step is to decode $\mathbf{m}$ from $\mathbf{b}$ by rounding and modular reduction. Notably, the decryption, and in particular, the second step, can succeed if and only if the error has a bounded value, i.e., $\norm{\mathbf{e}_{\mathsf{CT}}}<\Delta/2$.

\subsection{(Public-Key) Threshold FHE}
\label{sec:pre-mphe}

Traditional FHE is typically used in the single-server-single-client architecture. The client is the owner of private data and the server is used to evaluate some function over the ciphertexts of private data. In this architecture, the client owns the secret key, along with the full authority to decrypt all data. Nonetheless, in some scenario, the data may come from multiple clients and the decryption authority should be distributed among them.
Threshold fully homomorphic encryption (ThFHE) is proposed for such a scenario. Particularly, almost all ThFHE schemes require public-key encryption, as the clients are not allowed to own the secret key and can only use public key for encryption.
Although ThFHE has the same pipeline as traditional FHE (i.e., secret/public/evaluation key generation, encryption, homomorphic evaluation, and decryption), these operations are modified as follows to distribute the decryption authority.

For \textit{key generation}, ThFHE asks each party to sample a local secret key and the (implicit) global secret key is the summation of all local secret keys.
In addition, to support \textit{public-key} encryption and evaluation, all parties need to cooperate to generate global public/evaluation keys from their local secret keys.
To this end, the key generation algorithms are modified to multi-party protocols. 
These key generation protocols are executed only once in the setup stage.
Once these global public encryption and evaluation keys are generated, they are fixed and will be reused for encryption and evaluation in the subsequent computation stage. Since then, we follow the setting of existing work~\cite{Threshold,systemImplementation} and assume that all parties remain available in the setup stage.
For \textit{decryption}, ThFHE also modifies it to a multi-party protocol. This protocol involves a threshold parameter $T$ ($T\leq N$): the decryption can succeed if and only if at least $T$ parties participate.
We call such a scheme $T$-out-of-$N$ ThFHE. Correspondingly, the ThFHE schemes whose decryption relies on the participation of \textit{all} parties are called \textit{full}-threshold schemes with $T=N$.

We note that the decryption protocol can be re-executed to decrypt multiple ciphertexts. When decrypting, the parties who participate in the key generation protocol may fail to participate in decryption. Hence, we prefer those ThFHE schemes who can successfully decrypt even if only \textit{part} of parties participate. These preferred schemes should allow the elastic choice of $T$ according to the system's characteristics. Particularly, if there are at most $N^\prime$ non-participants in the system, then the setting of $T=N-N^\prime$ can support to decrypt. We call such a scheme arbitrary-threshold FHE (AThFHE). How to design AThFHE, and, in particular, its decryption protocol, is the core problem studied in this paper.

\section{ThFHE via ApproxSS}

\label{sec:thmphe}

In this section, we formulate approximate secret sharing (ApproxSS) in Section \ref{sec:ass-def} and establish a reduction from ThFHE construction to the design of ApproxSS in Section \ref{sec:thmphe-2}.

\subsection{Formulation of ApproxSS}
\label{sec:ass-def}
Below we formalize a novel primitive called approximate secret sharing (ApproxSS). 
Compared with vanilla SS, a $T$-out-of-$N$ ApproxSS has an additional operation called \emph{approximate recovery}. Informally, this operation recovers an approximate message from any $T$ shares of the message, such that 1) the approximate message is ``close'' to the original message and 2) the adversary corrupting $T-1$ parties can not learn ``too much information'' about the original message via this operation.
To formulate this operation, a direct way is to restrict it as a one-shot algorithm like $\mathtt{Share}$ and $\mathtt{Rec}$ in vanilla SS. Still, to thoroughly explore all possible constructions, we allow this operation to take multiple rounds, during which parties can interact with each other.
Hence, we formulate the approximate recovery as a multi-party protocol and define ApproxSS as follows.

\begin{definition}[Approximate Secret Sharing]

Given a message space $\mathcal{M}$, a $T$-out-of-$N$ approximate secret sharing scheme is a pair $(\mathtt{Share},\Pi_\mathtt{ApproxRec})$ defined as follows:

\noindent\textbf{Share}: $\{s_i\}_{i\in[N]}$$\gets$$\mathtt{Share}(m)$. On input a message $m\in\mathcal{M}$, this algorithm outputs $N$ shares $\{s_i\}_{i\in[N]}$;

\noindent\textbf{Approximate Recovery}: $m^\prime\gets\Pi_\mathtt{ApproxRec}(\{s_i\}_{i\in\mathcal{T}},\chi)$. On input shares $\{s_i\}_{i\in \mathcal{T}}$ from a set $\mathcal{T}$ and a distribution $\chi$ on $\mathcal{M}$, this multi-party protocol outputs a common approximate message $m^\prime\in \mathcal{M}$ to all parties.
\end{definition}

In the above definition, the distribution $\chi$ is used to characterize how much information about $m$ is allowed to be leaked via the execution of approximate recovery.
In other words, an adversary who corrupts $T-1$ parties cannot learn anything about $m$ other than a random value $m_\chi=m+x$ with $x\sample \chi$ and what is implied by this value.
For example, if $m=8$ and $\chi$ is a uniform distribution over $\{-1,0,1\}$, then the adversary could learn $m_\chi=9$ and further infer $m=8,9,10$ with equal probability, but nothing else. We term this property of ApproxSS \textit{approximate security}.
To formulate this property, we introduce two experiments $\mathsf{Expt}_{\mathcal{A},\mathsf{Real}}(\chi)$ and $\mathsf{Expt}_{\mathcal{A},\mathsf{Ideal}}(\chi)$, as described in Figure \ref{fig:ideal-world}. Particularly, the $\mathsf{Expt}_{\mathcal{A},\mathsf{Ideal}}(\chi)$ is the execution of ApproxSS over message $m$ and distribution $\chi$. In contrast, the view of adversary in $\mathsf{Expt}_{\mathcal{A},\mathsf{Ideal}}(\chi)$ is simulated from either public parameters (i.e., $N,T,\mathcal{M},\mathcal{T}_r$), the internal states of corrupted parties (i.e., $\{s_i\}_{i\in\mathcal{N}_A}$), or a random value $m_\chi=m+x$ with $x\sample\chi$. Hence, if the adversary fails to distinguish between two experiments, then it can learn nothing about $m$ except that $m$ can be a random value $m_\chi-x$ with $x\sample \chi$, implying the approximate security of ApproxSS.
We formalize the above discussion as follows.

\begin{definition}[Approximate Security]
\label{def:approx-security}
For an ApproxSS whose $\Pi_\mathtt{ApproxRec}$ consists of $R$ rounds, it satisfies $\chi$-approximate security if the following claim holds: There exists $R+1$ PPT simulator algorithms $(S_0,S_1,\ldots,S_R)$ such that for any PPT adversary $\mathcal{A}$, the experiments $\mathsf{Expt}_{\mathcal{A},\mathsf{Real}}(\chi)$ and $\mathsf{Expt}_{\mathcal{A},\mathsf{Ideal}}(\chi)$ in Figure \ref{fig:ideal-world} are indistinguishable.
\end{definition}
\begin{figure}[h]
    \centering
    \begin{center}
    \fbox{%
    \procedure{$\mathsf{Expt}_{\mathcal{A},\mathsf{Real}}(\chi)$: Real-World Experiment}{%
\pcln \mathcal{A}\text{ outputs $N,T,\mathcal{M}$, a message $m\in\mathcal{M}$, and a set $\mathcal{N}_A$ of}\\
\text{$T-1$ corrupted clients}\\
\pcln \text{The challenger runs \underline{\color{red}$\{s_i\}_{i\in[N]}\gets\mathtt{Share}(m)$} and sends}\\
\text{$\{s_i\}_{i\in\mathcal{N}_A}$ to }\mathcal{A}\\
\pcln \text{The $\Pi_\mathtt{ApproxRec}$ protocol is performed as follows:}\\
\text{In each round $r$, $\mathcal{A}$ selects a set $\mathcal{T}_r$ consisting of $T$ participants,}\\
\text{then the challenger \underline{\color{red}runs the $r$-th round of the protocol for parties}}\\
\text{\underline{\color{red}in $\mathcal{T}_r$} and provides the transcript $\mathsf{Trans}_r$ that can be observed by}\\
\text{parties in $\mathcal{N}_A$ to $\mathcal{A}$}\\
\pcln \text{At the end, $\mathcal{A}$ outputs a distinguishing bit }b
}
    }
  
    \fbox{%
    \procedure{$\mathsf{Expt}_{\mathcal{A},\mathsf{Ideal}}(\chi)$: Ideal-World Experiment}{%
\pcln \mathcal{A}\text{ outputs $N,T,\mathcal{M}$, a message $m\in\mathcal{M}$, and a set $\mathcal{N}_A$ of}\\
\text{$T-1$ corrupted clients}\\
\pcln \text{The challenger runs \underline{\color{red}$\{s_i\}_{i\in\mathcal{N}_A}\gets S_0(N,T,\mathcal{M})$} and sends}\\
\text{$\{s_i\}_{i\in\mathcal{N}_A}$ to }\mathcal{A}\\
\pcln \text{The challenger \underline{\color{red}samples $m_\chi\gets m+x$ with $x\sample\chi$} and then}\\
\text{simulates the $\Pi_\mathtt{ApproxRec}$ protocol as follows:}\\
\text{In each round $r$, $\mathcal{A}$ selects a set $\mathcal{T}_r$ consisting of $T$ participants,}\\
\text{then \underline{\color{red}the challenger generates transcript $\mathsf{Trans}_r^\prime$ by running}}\\
\text{\underline{\color{red}$\mathsf{Trans}_r^\prime\gets S_r(m_\chi,\{s_i\}_{i\in\mathcal{N}_A},\mathcal{T}_r)$} and sends $\mathsf{Trans}_r^\prime$ to }\mathcal{A}\\
\pcln \text{At the end, $\mathcal{A}$ outputs a distinguishing bit }b^\prime
}
    }
\end{center}
    \caption{Description of $\mathsf{Expt}_{\mathcal{A},\mathsf{Real}}(\chi)$ and $\mathsf{Expt}_{\mathcal{A},\mathsf{Ideal}}(\chi)$. Their differences are highlighted by red, underlined parts.}
    \label{fig:ideal-world}
\end{figure}

We note that there exists a trivial solution for approximate recovery construction: a party directly outputs the approximate message as a uniformly-random value over $\mathcal{M}$. In this way, no information about message $m$ is leaked. Nonetheless, this construction is also somewhat meaningless. When applying ApproxSS in the real-world applications (e.g., the ThFHE construction in this work), we usually need the approximate message to be ``close'' to the original message. This property is called \textit{approximate correctness}. To formulate this property, we use a subset $\mathcal{M}_B\subsetneqq\mathcal{M}$ to denote the range of allowed difference between approximate message $m^\prime$ and original message $m$. The common setting of $\mathcal{M}_B$ contains the elements whose value is less than or equal to a given number $B$. For example, if $m=8$ and $B=1$, then $\mathcal{M}_B=\{-1,0,1\}$ and $m^\prime$ should belong to $\{7,8,9\}$. With the notion of $\mathcal{M}_B$, the approximate correctness is formally defined as follows.

\begin{definition}[Approximate Correctness]
\label{def:approxcorrect}
An ApproxSS satisfies $\mathcal{M}_B$-approximate correctness if and only if for the shares of any message $\{s_1,\ldots,s_N\}\gets\mathtt{Share}(m)$ and any set $\mathcal{T}$ of at least $T$ parties, the approximate message $m^\prime \gets \Pi_\mathtt{ApproxRec}(\{s_i\}_{i\in\mathcal{T}},\chi)$ satisfies $m^\prime-m\in\mathcal{M}_B$.
\end{definition}

Notably, the support\footnote{Roughly speaking, for a random variable, the support of its distribution is the set of its possible values with non-zero probability. In this paper, we mainly consider a discrete random variable $x\in\mathcal{M}$. In this case, the support of its distribution $\chi$ is defined as the set $\{\Tilde{x}\in\mathcal{M}\mid P(x=\Tilde{x})>0\}$.} of distribution $\chi$ is not necessarily the same as the set $\mathcal{M}_B$. For example, when $\chi$ is the distribution over $\{0\}$ with $P(x=0)=1$, we can construct an ApproxSS scheme whose $\Pi_{\mathsf{ApproxRec}}$ protocol outputs $m+1$ by modifying a vanilla SS scheme:
after the aggregator recovers $m$ via the recovery algorithm, it randomly outputs $m$, $m+1$, and $m-1$ instead of $m$. This protocol can only satisfies $\mathcal{M}_B$-approximate security with $\mathcal{M}_B=\{-1,0,1\}$, while this $\mathcal{M}_B$ is not equal to the support of $\chi$ (i.e., $\{0\}$).
Nonetheless, we also note that there could exist some relations between the support of $\chi$ and the set $\mathcal{M}_B$. For example, when $\mathcal{M}_B=\{0\}$, the approximate recovery protocol must output $m$. In this case, the adversary can learn the exact value of $m$ and the best we can expect for $\chi$-approximate security is $\chi$ over $\{0\}$ with $P(x=0)=1$. 
Since this paper focuses on the ApproxSS's efficient construction and application for ThFHE, we left the studies on relations between $\chi$ and $\mathcal{M}_B$ as an interesting problem for future research.

\subsection{ApproxSS-Based ThFHE Construction}
\label{sec:thmphe-2}

Next, we show that a $T$-out-of-$N$ ThFHE scheme can be constructed based on a $T$-out-of-$N$ ApproxSS scheme.
Such a construction includes two modifications on existing full-threshold FHE schemes. One is to add a step called secret key sharing to the key generation protocol and the other is to re-devise the decryption protocol. Below we present these two modifications and illustrate them in Figure \ref{fig:MPHE}.


\noindent {\bf Key Generation Protocol.}
In this work, we consider the \textit{public-key} ThFHE construction, in which the key generation protocol consists of two parts: local secret key sampling and global public key generation.
For local secret key sampling, each party $i$ produces its secret key $\mathsf{sk}_i\gets\mathtt{BFV.SKGen}()$.
For global public key generation, parties collaborate to generate a global public/evaluation key. These keys are published to all parties and will be used for encryption/evaluation later.

Recall that we assume all parties are available for key generation. Under this setting, we can employ the key generation protocol of existing full-threshold schemes in our construction, without requiring many modifications. In these schemes, by well-crafted design, the global public keys are almost equivalent to the output of single-party FHE's key generation algorithm, with the corresponding global secret key $\mathsf{sk}$ being the summation of local secret keys from \textit{all} parties, i.e., $\mathsf{sk}=\sum_{i\in[N]} \mathsf{sk}_i$. Under this design, the encryption and evaluation algorithms are just the same as those of the single-party FHE.
Meanwhile, the decryption protocol will rely on the local secret keys from all parties. If a local secret key is only known by its owner (say party $i$), then the decryption protocol can not decrypt once party $i$ does not participate. To avoid this situation, we add a new step called $\mathtt{SKShare}$ as the third part of key generation protocol.


As specified in Figure \ref{fig:MPHE}, this step is executed by every party $i$ after its local secret key $\mathsf{sk}_i$ is generated.
Through $\mathtt{SKShare}$, party $i$ invokes the $\mathsf{Share}$ algorithms of the ApproxSS scheme to generate $T$-out-of-$N$ shares of its local secret key $\mathsf{sk}_i$ (Line 1).
We particularly note that the message space of ApproxSS should be the ciphertext space of BFV, as the secret key lies in this space.
Then the party $i$ sends its generated shares to other parties and receives the shares of local secret keys from other parties (Line 2-4). By the linearity of ApproxSS, the summation of the $i$-th share of local secret keys is the $i$-th share of the summation of local secret keys.
Recall that the global secret key is exactly the summation of local secret keys. By summing all received shares up, party $i$ generates its $T$-out-of-$N$ share of global secret key (Line 5-6).
Notably, any $T$ shares contain enough information of global secret key and can enable the decryption to succeed. Therefore, through this secret key sharing step, the decryption protocol will not require all parties to participate.

\noindent {\bf Decryption Protocol.} 
The decryption protocol can be used to decrypt a long message from multiple ciphertexts. To highlight our key idea, we only discuss the case with one ciphertext $\mathsf{CT}=(\mathbf{c}_0,\mathbf{c}_1)$ and put the full description with multiple ciphertexts in Figure \ref{fig:MPHE}. Besides the ciphertext, the decryption protocol needs two public parameters as input: the BFV parameter $\Delta$ and the norm bound of smudging noises $B_{sm}$.
For ease of description, we decompose the decryption into three successive phases. Phases 1 and 3 are locally executed by every party and the aggregator, respectively, while Phase 2 is an interactive protocol between them.

\begin{figure}[htbp]
    \centering
    \begin{center}
    \fbox{%
    \procedure{\textbf{ThFHE Construction via ApproxSS}}{%
\mathtt{SKShare}(\mathsf{sk}_i)\text{// Executed after }\mathsf{sk}_i\gets \mathtt{BFV.SKGen}()\\
\pcln \{\mathsf{skShare}_{j,i}\}_{j\in[N]}\gets\mathtt{ApproxSS.Share}(\mathsf{sk}_i)\\
\pcln \textbf{For $j=1,\ldots,N$, do}\\
\pcln \qquad\text{Send }\mathsf{skShare}_{j,i}\text{ to party }j\\
\pcln \qquad\text{Receive }\mathsf{skShare}_{i,j}\text{ from party }j\\
\pcln \mathsf{skShare}_i\gets\sum_{j\in[N]}\mathsf{skShare}_{j,i}\\
\pcln \text{Output }\mathsf{skShare}_{i}
\pclb
\pcintertext[dotted]{Decryption Protocol $\Pi_{\mathsf{Dec}}$}
\textbf{Private Input of Party }i\textbf{:}\text{ global secret key share }\mathsf{skShare}_{i}\\
\textbf{Public Input: }\text{$C$ ciphertexts and parameters $\Delta,B_{sm}$}\\
    \textbf{Aggregator Output: } \text{$C$ plaintexts by decrypting ciphertexts}\\
\\
\textbf{Phase 1}\text{ // Executed by Party $i\in\mathcal{T}$}\\
\pcln \textbf{For the $c$-th ciphertext $\mathsf{CT}_c=(\mathbf{c}_{c,0},\mathbf{c}_{c,1})$, do}\\
\pcln \qquad\mathbf{b}_{c,i}\gets \mathbf{c}_{c,1}\cdot \mathsf{skShare}_{i}+ \mathbf{c}_{c,0}\\
\pcln \text{Concatenate $\{\mathbf{b}_{1,i},\ldots,\mathbf{b}_{C,i}\}$ as $\mathbf{b}_{i}$}\\
\textbf{Phase 2}\text{ // Interaction between parties and the aggregator}\\
\pcln \chi\gets\mathsf{Uniform}(  \{\mathbf{n}\mid \norm{\mathbf{n}}\leq B_{sm}\}) \\
\pcln \mathbf{b}^\prime\gets\mathtt{ApproxSS}.\Pi_{\mathsf{ApproxRec}}(\{\mathbf{b}_i\}_{i\in\mathcal{T}},\chi)\\
\pcln \text{Parse $\mathbf{b}^\prime$ as $\{\mathbf{b}^\prime_1,\ldots,\mathbf{b}^\prime_C\}$}\\
\textbf{Phase 3}\text{ // Executed by the aggregator}\\
\pcln \textbf{For the $c$-th ciphertext $\mathsf{CT}_c=(\mathbf{c}_{c,0},\mathbf{c}_{c,1})$, do}\\
\pcln \qquad\mathbf{m}_{c}^\prime\gets\mathbf{b}_{c}^\prime+\mathbf{c}_{c,0}\\
\pcln \qquad\text{Decode $\mathbf{m}_c\in\mathcal{R}_{M,P}$ from $\mathbf{m}_c^\prime\in\mathcal{R}_{M,Q}$}\\
\pcln \text{Output }\mathbf{m}_1,\ldots,\mathbf{m}_C
}
    }
\end{center}
    \caption{Illustration of Our ThFHE construction}
    \label{fig:MPHE}
\end{figure}

\noindent\textbf{Phase 1: Computation of Decryption Share.} Recall that by the $\mathtt{SKShare}$ step in key generation protocol, each party $i$ has learned a $T$-out-of-$N$ share $\mathsf{skShare}_{i}$ of global secret key $\mathsf{sk}$. With the ciphertext $\mathsf{CT}=(\mathbf{c}_0,\mathbf{c}_1)$, the knowledge of this share enables party $i$ to compute $\mathbf{b}_i=\mathbf{c}_0+\mathbf{c}_1\cdot \mathsf{skShare}_{i}$.
By the linearity of ApproxSS, $\mathbf{b}_i$ is a $T$-out-of-$N$ share of $\mathbf{b}= \mathbf{c}_0+\mathbf{c}_1\cdot \mathsf{sk}$. Since $\mathbf{b}$ is a desired value for BFV decryption, we call $\mathbf{b}_i$ the decryption share of party $i$.

\noindent\textbf{Phase 2: Recovery of Approximate Decryption.} A natural idea for decryption is to recover $\mathbf{b}$ from its shares computed in Phase 1 and further decode message $\mathbf{m}$ from $\mathbf{b}$. 
However, such a recovery process may result in the leakage of secret key:
since $\mathbf{b}=\Delta\cdot \mathbf{m}+\mathbf{e}_{\mathsf{CT}}$, the knowledge of $\mathbf{b}$ and decryption result $\mathbf{m}$ will leak the ciphertext noise $\mathbf{e}_{\mathsf{CT}}$, which helps to further reveal the secret key.
To avoid the leakage, existing works~\cite{mpcviaFHE0,mpcviaFHE} propose \textit{noise smudging} technique to ``smudges out'' any small noise $\mathbf{e}_{\mathsf{CT}}$ by adding a large noise.
Formally, they prove the following lemma.
\begin{lemma}[Noise Smudging]
Let $B_1$ and $B_2$ be positive integers, and let $e_1\in[-B_1,B_1]$ be a fixed integer. Let $e_2\sample[-B_2,B_2]$ be chosen uniformly at random. Then the distribution of $e_2$ is statistically indistinguishable from that of $e_2+e_1$ as long as $B_1/B_2\in\mathsf{negl}$. 
\end{lemma}

Suppose that the error in ciphertext $\mathsf{CT}$ is bounded by $B_{\mathsf{CT}}$, i.e., $\norm{\mathbf{e}_{\mathsf{CT}}}\leq B_{\mathsf{CT}}$. Then for some ``sufficiently large'' $B_{sm}$ with $B_{\mathsf{CT}}/B_{sm}\in\mathsf{negl}$, we can use this lemma to prove that the value of $\mathbf{b}+\mathbf{e}_{sm}=\Delta\cdot \mathbf{m}+\mathbf{e}_{\mathsf{CT}}+\mathbf{e}_{sm}$ can be simulated as $\Delta\cdot \mathbf{m}+\mathbf{e}_{sm}$, where $\mathbf{e}_{sm}\sample[-B_{sm},B_{sm}]$.
This further indicates that the value of $\mathbf{b}+\mathbf{e}_{sm}$ is safe to be learned by the adversary, as the adversary can also generates the $\Delta\cdot \mathbf{m}+\mathbf{e}_{sm}$ by itself without using any secret information.

Inspired by this technique, we ask parties and the aggregator to conduct the approximate recovery protocol, with its input $\chi$ being the uniform distribution over $[-B_{sm},B_{sm}]$.
Based on the approximate security of ApproxSS, we can guarantee that the adversary cannot learn anything about $\mathbf{b}$ via the execution of approximate recovery protocol, except a random value $\mathbf{b}+\mathbf{e}_{sm}$ that is already safe to be disclosed according to the noise smudging technique. Therefore, we can further prove the security of our ThFHE construction.

Notably, researchers~\cite{realworldApp,peter,moduluspolynomial} further show that the noise smudging technique still works well with the $B_{sm}$ in the polynomial order rather than $B_{\mathsf{CT}}/B_{sm}\in\mathsf{negl}$.
We note that our construction on ThFHE as well as the ApproxSS scheme is compatible with this new conclusion. In fact, we just need to switch the parameter $B_{sm}$ from exponential order to polynomial order, while the other parts can remain unchangeable.

\noindent\textbf{Phase 3: Output of Final Plaintext.}
At the end of Phase 2, the aggregator learn an approximate message $\mathbf{b}^\prime$ that is close to the original message $\mathbf{b}$. Suppose that the ApproxSS scheme satisfies the $\mathcal{M}_B$-approximate correctness for some $\mathcal{M}_B=[-B,B]$, then the norm of total error in $\mathbf{b}^\prime$ is bounded by $B_{\mathsf{CT}}+B$. As long as $B<\Delta/2-B_{\mathsf{CT}}$, the aggregator can successfully decode plaintext $\mathbf{m}$ from $\mathbf{b}^\prime$.


The following theorems state the correctness and security of our ThFHE construction. Although the description of Phases 1 -- 3 has provided some reasoning, we formally prove them in Appendix \ref{appendix-proof-1} and Appendix \ref{appendix-proof-2}, respectively.

\begin{theorem}[Correctness of ThFHE]
\label{theo:MPHE-correctness}
Given a $T$-out-of-$N$ ApproxSS scheme with linearity and $\mathcal{M}_B$-approximate correctness, the ThFHE construction in Figure \ref{fig:MPHE} is a correct $T$-out-of-$N$ scheme with guaranteed output delivery if $B<\Delta/2-B_{\mathsf{CT}}$. 
\end{theorem}
\begin{theorem}[Security of ThFHE]
\label{theo:MPHE-security}
Given a $T$-out-of-$N$ ApproxSS scheme with vanilla security and $\chi$-approximate security, the ThFHE construction in Figure \ref{fig:MPHE} satisfies simulation-based security if $B_\mathsf{CT}/B_{sm}\in\mathsf{negl}$.
    
\end{theorem}


In addition, we note that the decryption protocol consists of the approximate recovery protocol (in Phase 2) and simple operations (in Phases 1 and 3) with complexity $\mathcal{O}(K)$, which is linear with respect to the length of ciphertext $K$ but keeps constant with respect to the number of parties $N$. Hence, the efficiency of decryption protocol is dominated by the efficiency of approximate recovery protocol, unless the approximate recovery protocol has less complexity than $\mathcal{O}(K)$, which is unlikely to happen. This inspires us to design efficient ApproxSS, especially under arbitrary threshold.

\section{ATASSES: Construction and Analysis}
\label{sec:ass}

This section starts with the challenges of ApproxSS constructions and the shortcomings of existing solutions in Section \ref{sec:ass-existing} and then presents how we address them in Section \ref{sec:ass-our}.

\subsection{Challenges and Existing Solutions}
\label{sec:ass-existing}

Existing works design several arbitrary-threshold (ATh)-ApproxSS schemes based on two types of linear SS. One is linear SS with small recovery coefficients, and the other is Shamir SS. Below we describe the challenges of ATh-ApproxSS design under these two types of SS and highlight the shortcomings of existing solutions.


\noindent {\bf ApproxSS with Small Recovery Coefficients.} Recall that for $T$-out-of-$N$ linear SS, each share consists of $L$ pieces and the message $m$ can be recovered as a linear combination $m=\sum_{i\in\mathcal{T},l\in[L]} w^{(\mathcal{T})}_{i,l}\cdot s_{i,l}$,
where $\mathcal{T}$ is a set of at least $T$ parties, $s_{i,l}$ is the $l$-th piece of the share of party $i$, and $w^{(\mathcal{T})}_{i,l}$ is the recovery coefficients. An existing ApproxSS adopts linear SS with $w^{(\mathcal{T})}_{i,l}$ being a small value like $0,1$, a.k.a, $\{0,1\}$-linear SS~\cite{Threshold}.


This type of SS can be easily boosted to ApproxSS and is thus friendly to ThFHE design. To do so, the approximate recovery protocol only needs one round. In this round, every party $i$ adds each of its piece $s_{i,l}$ ($l=1,\ldots,L$) a noise $n_{i,l}\sample\chi_B$. After collecting these noisy shares from any $T$ parties, the aggregator can recover the approximate message as follows
\begin{equation}
    m^\prime=\sum_{i\in\mathcal{T},l\in[L]} w^{(\mathcal{T})}_{i,l}\cdot (s_{i,l}+n_{i,l})=m+\sum_{i\in\mathcal{T},l\in[L]} w^{(\mathcal{T})}_{i,l}\cdot n_{i,l}.
\end{equation}
Since $w^{(\mathcal{T})}_{i,l}$ and $n_{i,l}$ both have bounded values, the difference between original message and approximate message also has a bounded value. Readers can easily check its approximate correctness and prove its approximate security.

However, this type of SS suffers from an extremely large size of each share in the arbitrary-threshold case. For example, in a $T$-out-of-$N$ replicated SS, each share consists of $\tbinom{N-1}{T-1}$ pieces, which may increase exponentially with the number of parties $N$. Hence, this type of ApproxSS and its induced ThFHE can only be used in some special cases with $\tbinom{N-1}{T-1}$ being a small value, such as in the full-threshold case ($N=T$) or in a small system (e.g., $N=2$ or $N=3$)~\cite{nuoyafangzhou}.

\noindent {\bf Shamir ApproxSS.} Shamir SS enjoys an advantage that its share has the same size as the message. Suppose that each party $i$ owns a share $s_i$, then the message $m$ is recovered as $m=\sum_{i\in\mathcal{T}} L^{(\mathcal{T})}_i\cdot s_i$,
where $\mathcal{T}$ can be any set of at least $T$ parties and $L^{(\mathcal{T})}_i$s are the Lagrange coefficients.

Although Shamir SS has the efficiency advantage, the Lagrange coefficients have two properties that make the design of Shamir ApproxSS very challenging.
One is unbounded value, i.e., the Lagrange coefficients can be arbitrarily large in the message space.
Another is unpredictablity. Lagrange coefficients are associated with the set $\mathcal{T}$ and their values can not be learned until the set $\mathcal{T}$ is determined.
These two features imply that Lagrange coefficients can be arbitrarily-large unknown numbers.
If each party $i$ adds a noise $n_i$ to its share $s_i$ for approximate security, then the recovered approximate message will be
\begin{equation}
    m^\prime=\sum_{i\in\mathcal{T}} L^{(\mathcal{T})}_i\cdot (s_i+n_i)=m+\sum_{i\in\mathcal{T}} L^{(\mathcal{T})}_i\cdot n_i.
\end{equation}
The difference between original message and approximate message is $\sum_{i\in\mathcal{T}} L^{(\mathcal{T})}_i\cdot n_i$, which may be blown up by $L^{(\mathcal{T})}_i$s and fail the approximate correctness. To overcome the ``blown-up noise'' challenge, existing works propose three ideas.
\begin{itemize}
    \item \textit{Type-\Rmnum{1}.} One idea is to adopt the ``clearing out the denominators'' technique~\cite{clearing-out} to scale $n_i$ so that $L^{(\mathcal{T})}_i\cdot n_i$ has a value bounded by $(N!)^3$~\cite{Threshold}. However, this bound is extremely loose and grows rapidly with the number of parties $N$. Recall that the ApproxSS requires the bound of $\sum_{i\in\mathcal{T}} L_i\cdot n_i$ to be smaller than the modulus of message space ($\mathcal{M}_B\subsetneqq\mathcal{M}$). Hence, this idea results in a large message space of ApproxSS with modulus being $\mathcal{O}(N\cdot(N!)^3)$. This huge message space incurs high complexity and difficulty for parameter instantiation. 
    \item \textit{Type-\Rmnum{2}.} The second idea is to coordinate the noise $n_i$s from all parties so that $n_i$s are the shares of some bounded value $n$~\cite{interactiveTFHE}. In other words, we have $n=\sum_{i\in\mathcal{T}} L^{(\mathcal{T})}_i\cdot n_i$ for any set $\mathcal{T}$. By this way, the difference between original and approximate messages is in fact $n$, whose value is bounded. However, the noise coordination asks each party to share a random noise with the same length as message (say $K$) to all of other parties. This may result in relatively high cost, especially when the values of $N$ and $K$ are both large. 
    \item \textit{Type-\Rmnum{3}.} Another idea is to assume that the set of participants $\mathcal{T}$ can be known in advance~\cite{efficient,Helium}. In this way, the set $\mathcal{T}$ as well as its associated Lagrange coefficients, can be learned and be used to generate a $T$-out-of-$T$ decryption share. Then the design of approximate recovery protocol is reduced to its counterpart in the full-threshold case, which can be fairly easy and efficient. Nonetheless, the set $\mathcal{T}$ of participants can be determined by random factors or even the adversary in the real world. Hence, it is impractical to assume the knowledge of $\mathcal{T}$ in advance.
\end{itemize}
In a word, existing Shamir-based constructions are either inapplicable in the real world or inefficient with high complexity.

\subsection{ATASSES for Efficient AThFHE}
\label{sec:ass-our}

Faced with the above deficiencies, we use a novel idea called ``encrypted share'' and propose ATASSES, an Arbitrary-Threshold ApproxSS scheme based on Encrypted Share idea. Next, we first introduce this idea, then describe a concrete construction, and discuss its application for AThFHE at last.

\noindent {\bf Main Idea.} Our goal is to build a Shamir ApproxSS with lower complexity. Suppose that the shares $s_i$s have been generated from the message $m$ by Shamir secret sharing, i.e., $m=\sum_{i\in\mathcal{T}} L^{(\mathcal{T})}_i\cdot s_i$ for any set $\mathcal{T}$ of at least $T$ parties. The key problem is how to devise the approximate recovery protocol. We note that the shares can not be sent to the aggregator without any protection. Otherwise, the aggregator can exactly recover the message and break the approximate security.
Correspondingly, we rely on a novel idea to protect the shares.

\noindent\emph{Part 1: Share Protection.} Different from existing constructions who directly add some noise to the share, we employ BFV secret-key encryption to protect shares. To distinguish from ThFHE's parameters, we use the superscript to denote the parameters of BFV secret-key encryption here. For example, the moduli for the plaintext space and ciphertext space are denoted by $P^\prime$ and $Q^\prime$, respectively, and the degree of BFV's secret key is $M^\prime$.
In this case, one ciphertext can only encrypt a degree-$M^\prime$ polynomial.
Nonetheless, the share $s_i$ can be a long vector with length $K\gg M^\prime$.
In this case, party $i$ breaks $s_i$ down into $C^\prime$ sub-vectors with length $M^\prime$ so that $K\leq C^\prime M^\prime$ and encrypts each sub-vector, respectively. 
Formally, the $k$-th sub-vector $s_{i,k}$ is encrypted to $\mathsf{CTs}_{i,k}$ as
\begin{equation}
    \mathsf{CTs}_{i,k}=(\mathbf{a}_k\cdot \mathsf{ek}_{i,1}+\mathbf{e}_{i,1,k}+\Delta^\prime \cdot s_{i,k},-\mathbf{a}_k),\label{eq:idea-CT-1}
\end{equation}
where $\mathbf{a}_k$ is a uniformly-random polynomial, $\mathsf{ek}_{i,1}$ is the (secret) encryption key of party $i$, $\mathbf{e}_{i,1,k}$ is an error polynomial with norm bounded by $B^\prime$, and $\Delta^\prime=\lfloor Q^\prime/P^\prime\rfloor$.

\noindent\emph{Part 2: Message Recovery.} 
Our first observation is that BFV secret-key encryption is linearly-homomorphic in the key. Specifically, given $\mathsf{CTs}_{i,k}$s from multiple parties $i\in\mathcal{T}$ and Lagrange coefficients $\{L_i^{(\mathcal{T})}\}_{i\in\mathcal{T}}$, if $\mathsf{CTs}_{i,k}[1]=-\mathbf{a}_k$ for any $i$, the aggregator can compute a new ciphertext $\mathsf{CTs}_k$ as
\begin{equation}
\label{eq:idea-CT-1-sum}
    \mathsf{CTs}_k=(\sum_{i\in\mathcal{T}} L_i^{(\mathcal{T})}\cdot\mathsf{CTs}_{i,k}[0],-\mathbf{a}_k).
\end{equation}
Substituting Eq.\eqref{eq:idea-CT-1} into Eq.\eqref{eq:idea-CT-1-sum} leads to the following:
\begin{equation}
    \begin{split}
        \mathsf{CTs}_k[0]=&\sum_{i\in\mathcal{T}} L_i^{(\mathcal{T})}\cdot\mathsf{CTs}_{i,k}[0]= \mathbf{a}_k\cdot \sum_{i\in\mathcal{T}} (L_i^{(\mathcal{T})}\cdot\mathsf{ek}_{i,1})\\
    &+\sum_{i\in\mathcal{T}} L_i^{(\mathcal{T})}\cdot \mathbf{e}_{i,1,k}+\Delta^\prime \cdot (\sum_{i\in\mathcal{T}} L_i^{(\mathcal{T})}\cdot s_{i,k}). 
    \end{split}
    \label{eq:idea-CT-1-sum-2}
\end{equation}

By observing Eq.\eqref{eq:idea-CT-1-sum-2}, we can tell that $\mathsf{CTs}_k$ is the encryption of $\sum_{i\in\mathcal{T}} L_i^{(\mathcal{T})}\cdot s_{i,k}=m_k$, with the key being $L_i^{(\mathcal{T})}\cdot\mathsf{ek}_{i,1}$. 
In other words, the aggregator can decrypt the original message $m_{k}$ from $\mathsf{CTs}_k$ if it learns decryption key $\mathsf{dk}=\sum_{i\in\mathcal{T}} L_i^{(\mathcal{T})}\cdot\mathsf{ek}_{i,1}$.
Nonetheless, ApproxSS requires to recover an approximate message rather than the original message. To solve this problem, we simply ask each party $i$ to encrypt a $B_{sm}$-bounded noise $n_{i,k}$ using another key $\mathsf{ek}_{i,2}$ but the same $\mathbf{a}_k$, i.e.,
\begin{equation}
    \mathsf{CTn}_{i,k}=(\mathbf{a}_k\cdot \mathsf{ek}_{i,2}+\mathbf{e}_{i,2,k}+\Delta^\prime \cdot n_{i,k},-\mathbf{a}_k),\label{eq:idea-CT-2}
\end{equation}
By using the linearly key-homomorphic property again, the aggregator combines $\mathsf{CTn}_{i,k}$s and $\mathsf{CTs}_{i,k}$s into an overall ciphertext $\mathsf{CTall}_k$ as following.
\begin{equation}
\label{eq:idea-CT-2-sum}
    \mathsf{CTall}_k=(\sum_{i\in\mathcal{T}} L_i^{(\mathcal{T})}\cdot\mathsf{CTs}_{i,k}[0]+\mathsf{CTn}_{i,k}[0],-\mathbf{a}_k).
\end{equation}
With the similar reasoning process, we can come to the conclusion: $ \mathsf{CTall}_k$ is the encryption of $\sum_{i\in\mathcal{T}} (L_i^{(\mathcal{T})}\cdot s_{i,k}+n_{i,k})$, with decryption key being $\mathsf{dk}=\sum_{i\in\mathcal{T}} L_i^{(\mathcal{T})}\cdot\mathsf{ek}_{i,1}+\mathsf{ek}_{i,2}$.
Since $\sum_{i\in\mathcal{T}} (L_i^{(\mathcal{T})}\cdot s_{i,k}+n_{i,k})=m_k+\sum_{i\in\mathcal{T}} n_{i,k}$, the decryption of $\mathsf{CTall}_k$ with $\mathsf{dk}$ is exactly the desired approximate message. 
Recall that the value of $n_{i,k}$ is bounded by $B_{sm}$. The difference between original and approximate messages is bounded by $N\cdot B_{sm}$, which satisfies approximate correctness.

Before presenting how the aggregator learns $\mathsf{dk}$, we note that the decryption of overall ciphertext $\mathsf{CTall}_k$ may fail if its error exceeds $\Delta^\prime/2$. Nonetheless, we only need to select a slightly larger ciphertext space to fix this problem. Particularly, the error in $\mathsf{CTall}_k$ is $\sum_{i\in\mathcal{T}}(L^{(\mathcal{T})}_i\mathbf{e}_{i,1,k}+\mathbf{e}_{i,2,k})$. Since $L^{(\mathcal{T})}_i$s are in the message space of Shamir SS as $s_i$, which is also the plaintext space of BFV secret-key encryption, their values are bounded by $P^\prime-1$. Recall that the values of original errors $\mathbf{e}_{i,1,k}$ and $\mathbf{e}_{i,2,k}$ are bounded by $B^\prime$. Then the overall error is bounded by $N\cdot P^\prime\cdot B^\prime$. Hence, although Lagrange coefficients $L^{(\mathcal{T})}_i$s also blow the error up, if we set $N\cdot P^\prime\cdot B^\prime<\Delta^\prime/2$ or more strictly $Q^\prime>2(P^\prime)^2B^\prime N+2P^\prime=\mathcal{O}(N)$, the decryption will succeed to output correct approximate message. In other words, our scheme only requires $Q^\prime=\mathcal{O}(N)$ and has no requirement on $P^\prime$. In contrast, Type-\Rmnum{1} Shamir ApproxSS~\cite{Threshold} requires the message space with $P^\prime=\mathcal{O}(N\cdot (N!)^3)$ that is much larger than the ciphertext space of our scheme. This is one reason why our ``encrypted share'' idea performs better.
\begin{figure}[htbp]
    \centering
    \begin{center}
    \fbox{%
    \procedure{\textbf{ATASSES Construction}: $\Pi_{\mathsf{ApproxRec}}$}{%
\textbf{Private Input of Party }i\textbf{:}\text{ a share }s_i\text{ of message }m\\
\textbf{Public Input: }\chi = \mathsf{Uniform}(\{n\mid \norm{n}\leq B_{sm}\})\\
    \textbf{Aggregator Output: }\text{an approximate message }m^\prime\\
\\
\mathtt{PartyR1}(s_i)\\
\pcln \mathsf{ek}_{i,1}\gets\mathtt{BFV.SKGen}(),\ \mathsf{ek}_{i,2}\gets\mathtt{BFV.SKGen}()\\
\pcln \{\mathsf{ekShare}_{j,i,1}\}_{j\in[N]}\gets\mathtt{ShamirSS.Share}(\mathsf{ek}_{i,1})\\
\pcln \{\mathsf{ekShare}_{j,i,2}\}_{j\in[N]}\gets\mathtt{ShamirSS.Share}(\mathsf{ek}_{i,2})\\
\pcln \text{Sends }\mathsf{ekShare}_{j,i,1},\mathsf{ekShare}_{j,i,2}\text{ to party }j\\
\pcln \text{Breaks $s_i$ down into length-$M^\prime$ sub-vectors}\\
\pcln \textbf{For the $k$-th length-$M^\prime$ sub-vector $s_{i,k}$, do}\\
\pcln \qquad \mathbf{a}_k\sample \mathsf{CRS}, \quad n_{i,k}\sample \chi \\
\pcln \qquad \mathsf{CTs}_{i,k} \gets \mathtt{BFV.SKEnc}(\mathsf{ek}_{i,1},s_{i,k},\mathbf{a}_k)\\
\pcln \qquad\mathsf{CTn}_{i,k} \gets \mathtt{BFV.SKEnc}(\mathsf{ek}_{i,2},n_{i,k},\mathbf{a}_k)\\
\pcln \qquad\text{Sends }\mathsf{CTs}_{i,k},\mathsf{CTn}_{i,k}\text{ to the aggregator}\\
\mathtt{AggregatorR1}()\\
\text{// Executed after collecting $\mathsf{CTs}_{i,k},\mathsf{CTn}_{i,k}$ for all $k$s from set $\mathcal{T}$}\\
\pcln \text{Computes }\{L_i^{(\mathcal{T})}\}_{i\in\mathcal{T}} \text{ and sends }\{L_i^{(\mathcal{T})}\}_{i\in\mathcal{T}}\text{ to all parties}\\
\mathtt{PartyR2}(\{L_i^{(\mathcal{T})}\}_{i\in\mathcal{T}}, \{\mathsf{ekShare}_{i,j,1},\mathsf{ekShare}_{i,j,2}\}_{j\in[N]})\\
\pcln \mathsf{dkShare}_j = \sum_{i\in\mathcal{T}} (L_i^{(\mathcal{T})}\cdot \mathsf{ekShare}_{i,j,1}+\mathsf{ekShare}_{i,j,2})\\
\pcln \text{Sends }\mathsf{dkShare}_j\text{ to the aggregator}\\
\mathtt{AggregatorR2}(\{\mathsf{CTs}_{i,k},\mathsf{CTn}_{i,k}\}_{i\in\mathcal{T}},\{\mathsf{dkShare}_j\}_{j\in\mathcal{T}_2})\\
\text{// Executed after collecting $\mathsf{dkShare}_j$ from set $\mathcal{T}_2$}\\
\pcln \mathsf{dk} \gets \mathtt{ShamirSS.Rec}(\{\mathsf{dkShare}_j\}_{j\in\mathcal{T}_2})\\
\pcln \textbf{For the $k$-th ciphertext $(\mathsf{CTs}_{i,k},\mathsf{CTn}_{i,k})$, do}\\
\pcln \qquad \mathsf{CTall}_{k}\gets(\sum_{i\in\mathcal{T}} (L_i^{(\mathcal{T})}\cdot\mathsf{CTs}_{i,k}[0]+\mathsf{CTn}_{i,k}[0]),-\mathbf{a}_k)\\
\pcln \qquad m^\prime_k\gets\mathtt{BFV.Dec}(\mathsf{dk},\mathsf{CTall}_{k})\\
    \pcln \text{Concatenate all $m^\prime_k$s as $m^\prime$}
}
    }
\end{center}
    \caption{The approximate recovery protocol of ATASSES.}
    \label{fig:approxrec}
\end{figure}

\noindent\emph{Part 3: Secure Computation of Decryption Key.} Now, the only step left is to obtain the decryption key $\mathsf{dk}=\sum_{i\in\mathcal{T}} L_i^{(\mathcal{T})}\cdot\mathsf{ek}_{i,1}+\mathsf{ek}_{i,2}$. 
We first note that $\mathsf{dk}$ is safe to disclose: following the security proof in~\cite{acorn}, the ciphertexts are still indistinguishable from random values even when $\mathsf{dk}$ is revealed. Readers can find more details in Appendix \ref{appendix:proof-atasses}.
To see how to compute the decryption key, we recall that $L^{(\mathcal{T})}_i$s are associated with the set $\mathcal{T}$ and can NOT be revealed before the set $\mathcal{T}$ is known. 
Since $L^{(\mathcal{T})}_i$s are necessary for computing $\mathsf{dk}$, the aggregator needs an extra round after the set $\mathcal{T}$ is known.
Nonetheless, the party $i\in\mathcal{T}$ who encrypts $s_i$ and $n_i$ may not participate in this extra round. If so, no one knows its encryption keys $\mathsf{ek}_{i,1},\mathsf{ek}_{i,2}$ that are also necessary for the $\mathsf{dk}$ computation. 

Our solution to this problem consists of two rounds. In the first round, each participant $i\in\mathcal{T}$ not only encrypts $s_i$ and $n_i$, but also shares its encryption keys with other parties using $T$-out-of-$N$ Shamir SS.
At the end of the first round, the knowledge of the set $\mathcal{T}$ becomes available, as well as its associated $L^{(\mathcal{T})}_i$s.
With the knowledge of $L^{(\mathcal{T})}_i$s, in the second round, each party $j$ computes one decryption key share from its received shares of $\mathsf{ek}_{i,1}$ and $\mathsf{ek}_{i,2}$ by the linear property of Shamir SS. 
Suppose that the set of participants in the second round is $\mathcal{T}_2$.
After collecting the shares of $\mathsf{dk}$ from $\mathcal{T}_2$, the aggregator recovers $\mathsf{dk}$ and further obtains the approximate message by decrypting $\mathsf{CTall}_{k}$.
Here, the set of participants $\mathcal{T}$ and $\mathcal{T}_2$ can be different. Therefore, our scheme does not require to learn the set of participants in advance, which is a key difference from the Type-\Rmnum{3} Shamir ApproxSS~\cite{efficient}.

We note that our solution also asks each party $i$ to share its encryption keys with other parties, which leads to quadratic computation complexity $O(N^2)$ with respect to the number of parties $N$. Nonetheless, the length of encryption key is the intrinsic parameter of BFV, which remains constant with respect to the length of message.
Therefore, in our solution, the complexity of sharing does not grow with the message's length $K$. For comparison, the complexity of Type-\Rmnum{2} Shamir ApproxSS asks to share a noise whose length is the same as that of message. 
As a result, its complexity of sharing is $\mathcal{O}(N^2 K)$ and grows with message's length. This is another reason why our solution performs better.

\noindent {\bf Concrete Construction.} As an implementation of the above idea, the $\Pi_\mathsf{ApproxRec}$ protocol of ATASSES consists of two rounds, as illustrated in Figure \ref{fig:approxrec}.

\begin{itemize}[leftmargin=*]
     \item \textit{Round 1: Encryption of Share and Noise.} Each party $i$ executes algorithm $\mathtt{PartyR1}$ for two tasks. One is to generate and share two encryption keys with other parties (Line 1-4). The other is to encrypt its share $s_i$ and noise $n_i$ using BFV secret-key encryption and sends the ciphertexts to the aggregator (Line 5-10). 
    After collecting at least $T$ groups of ciphertexts of $s_i$ and $n_i$ from set $\mathcal{T}$, the aggregator executes $\mathtt{AggregatorR1}$ to compute the Lagrange coefficients associated with set $\mathcal{T}$ and send them to all parties. We note that the successful decryption of ATASSES relies on a condition, namely, the set of parties who send the ciphertexts should be identical to the set of parties who send the shares of encryption keys, which is both denoted by symbol $\mathcal{T}$. Our system model in Section 2.2 assumes that this condition is met for at least $T$ parties. In practice, we can take the aggregator as the communication relay between parties, so that the aggregator can monitor the party-to-party message delivery and guarantee to satisfy this condition.
    \item  \textit{Round 2: Decryption of Approximate Message.} Each party $j$ executes $\mathtt{PartyR2}$ to compute its decryption key share $\mathsf{dkShare}_{j}$. After collecting at least $T$ shares from set $\mathcal{T}_2$, the aggregator executes $\mathtt{AggregatorR2}$ to recover the decryption key $\mathsf{dk}$ from $\{\mathsf{dkShare}_{j}\}_{j\in\mathcal{T}_2}$ (Line 14) and further decrypt every overall ciphertexts $\mathsf{CTall}_k$s (Line 15-17). The concatenation of decryption results is the desired output $m^\prime$.

    
\end{itemize}

\noindent {\bf Application to ThFHE.}
To construct ThFHE based on ATASSES, the ciphertext space of ThFHE should be set as the message space of ATASSES, so that $b$ in ThFHE can be regarded as $m$ in ATASSES. In addition, as we will show later, the ATASSES can achieve $\mathcal{M}_B$-approximate correctness with $B=T\cdot B_{sm}$. Recall that the correctness of ThFHE requires $B<\Delta/2-B_{\mathsf{CT}}$. Hence, when applying ATASSES for ThFHE construction, we need to set $T\cdot B_{sm}<\Delta/2-B_{\mathsf{CT}}$.

\subsection{Performance Analysis of ATASSES}
\label{sec:ass-analysis}
We analyze ATASSES in terms of its correctness, security, and efficiency. The approximate correctness and security are given by Theorem \ref{theo:ATASSES}. The complexity of $\Pi_{\mathsf{ApproxRec}}$ protocol of ATASSES along with other ApproxSS schemes is listed in Table \ref{tab:complexity}. Due to the page limit, the proof and analysis details is deferred to Appendix \ref{appendix:proof-atasses} and Appendix \ref{appendix-proof-4}, respectively. We note that Type-\Rmnum{3} Shamir ApproxSS has a different assumption with other schemes, i.e., the knowledge of participant set is known in advance. Compared with existing schemes that do not rely on this assumption, ATASSES reduces the computation (resp. communication) complexity from $\mathcal{O}(N^2\cdot K)$ to $\mathcal{O}(N^2+NK)$ (resp. $\mathcal{O}(N\cdot K)$ to $\mathcal{O}(N+K)$). Although Type-\Rmnum{3} scheme shows lower complexity, it may lose superiority when the knowledge of participant set is fault or even unavailable. In this case, Type-\Rmnum{3} scheme can only randomly search from $\tbinom{N}{T}$ possibilities until finding the correct set $\mathcal{T}$, which may yield a significantly higher (average) cost than ATASSES.

\begin{theorem}[ATASSES's Properties]
\label{theo:ATASSES}
ATASSES satisfies $\chi$-approximate correctness and $\mathcal{M}_B$-approximate security under RLWE-hardness assumption for $\chi=\mathsf{Uniform}(\{n\mid \norm{n}\leq B_{sm}\})$ and $\mathcal{M}_B=\{n\mid \norm{n}\leq T\cdot B_{sm}\}$.
    
\end{theorem}

\begin{table}[]
\centering
\caption{Comparison between ATASSES and existing ApproxSS schemes in terms of $\Pi_{\mathsf{ApproxRec}}$ protocol's communication complexity (Comm.), computation complexity (Comp.), and round number.}
\label{tab:complexity}
\resizebox{\columnwidth}{!}{%
\begin{tabular}{|c|c|cccc|}
\hline
\multirow{2}{*}{} & \multirow{2}{*}{\begin{tabular}[c]{@{}c@{}}$\{0,1\}$-\\ ApproxSS\end{tabular}} & \multicolumn{4}{c|}{Shamir-based ApproxSS} \\ \cline{3-6} 
 &  & \multicolumn{1}{c|}{Type-\Rmnum{1}~\cite{Threshold}} & \multicolumn{1}{c|}{Type-\Rmnum{2}~\cite{interactiveTFHE}} & \multicolumn{1}{c|}{Type-\Rmnum{3}~\cite{efficient,Helium}} & \textbf{ATASSES} \\ \hline
Comm. & $\mathcal{O}(N^{4.2}K)$ & \multicolumn{1}{c|}{$\mathcal{O}(N K)$} & \multicolumn{1}{c|}{$\mathcal{O}(NK)$} & \multicolumn{1}{c|}{$\mathcal{O}(K)$} & $\mathcal{O}(N+K)$ \\ \hline
Comp. & $\mathcal{O}(N^{5.2}K)$ & \multicolumn{1}{c|}{$\mathcal{O}(N^2K)$} & \multicolumn{1}{c|}{$\mathcal{O}(N^2K)$} & \multicolumn{1}{c|}{$\mathcal{O}(NK)$} & $\mathcal{O}(N^2 +NK)$ \\ \hline
\begin{tabular}[c]{@{}c@{}}Round\\ Number\end{tabular} & 1 & \multicolumn{1}{c|}{1} & \multicolumn{1}{c|}{2} & \multicolumn{1}{c|}{1} & 2 \\ \hline
\end{tabular}%
}
\end{table}


\section{Experiment Evaluation}
\label{sec:experiment}
\begin{figure*}[htbp]
	\centering
	\begin{subfigure}{0.24\linewidth}
		\centering
		\includegraphics[width=\linewidth]{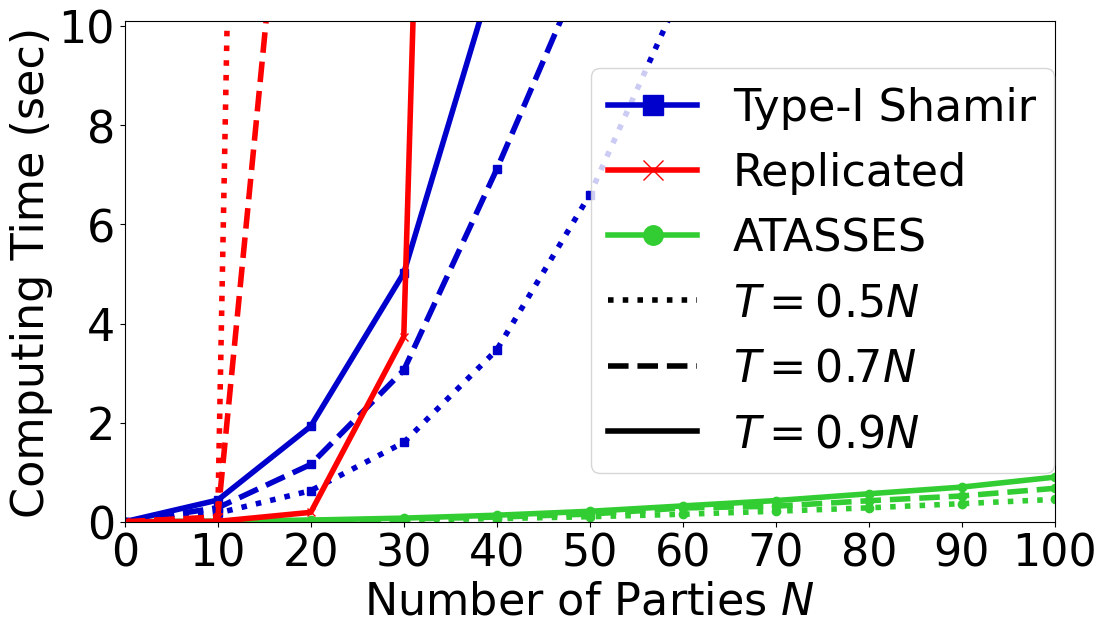}
		\caption{One-round ($K=5M^\prime$)}
		\label{time1k5}
	\end{subfigure}
	\centering
	\begin{subfigure}{0.24\linewidth}
		\centering
		\includegraphics[width=\linewidth]{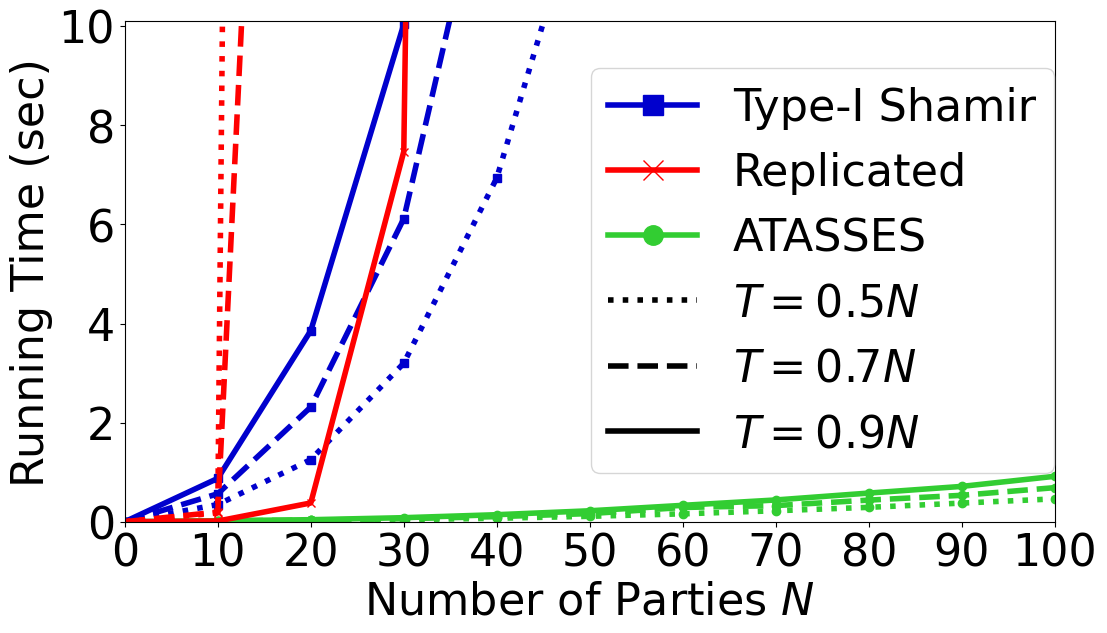}
		\caption{One-round ($K=10M^\prime$)}
		\label{time1k10}
	\end{subfigure}
	\centering
	\begin{subfigure}{0.24\linewidth}
		\centering
		\includegraphics[width=\linewidth]{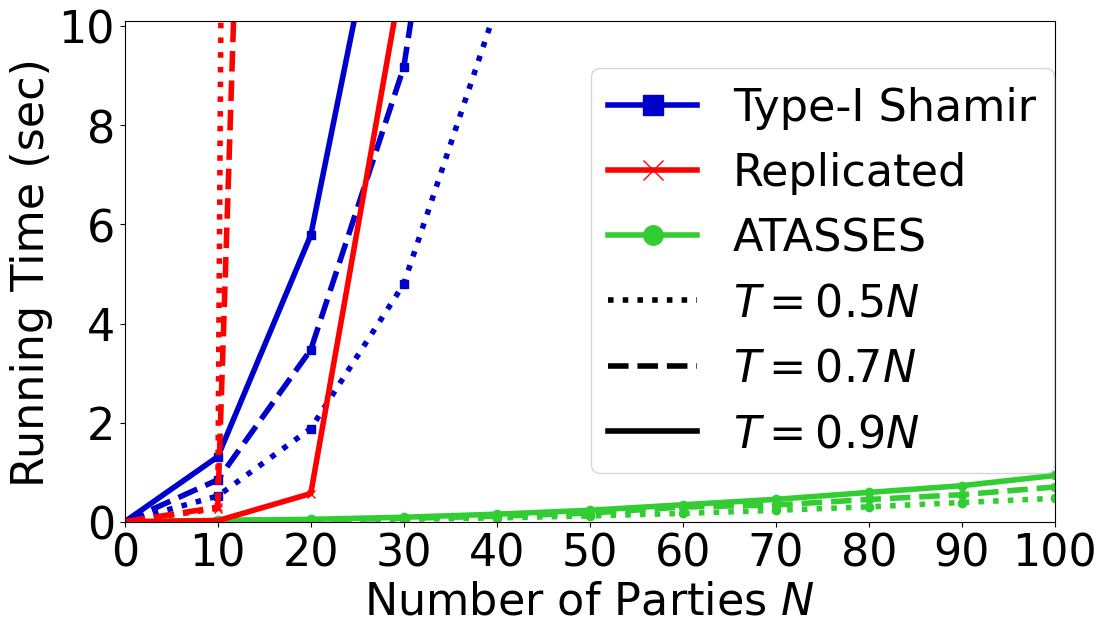}
		\caption{One-round ($K=15M^\prime$)}
		\label{time1k15}
	\end{subfigure}
        \centering
	\begin{subfigure}{0.24\linewidth}
		\centering
		\includegraphics[width=\linewidth]{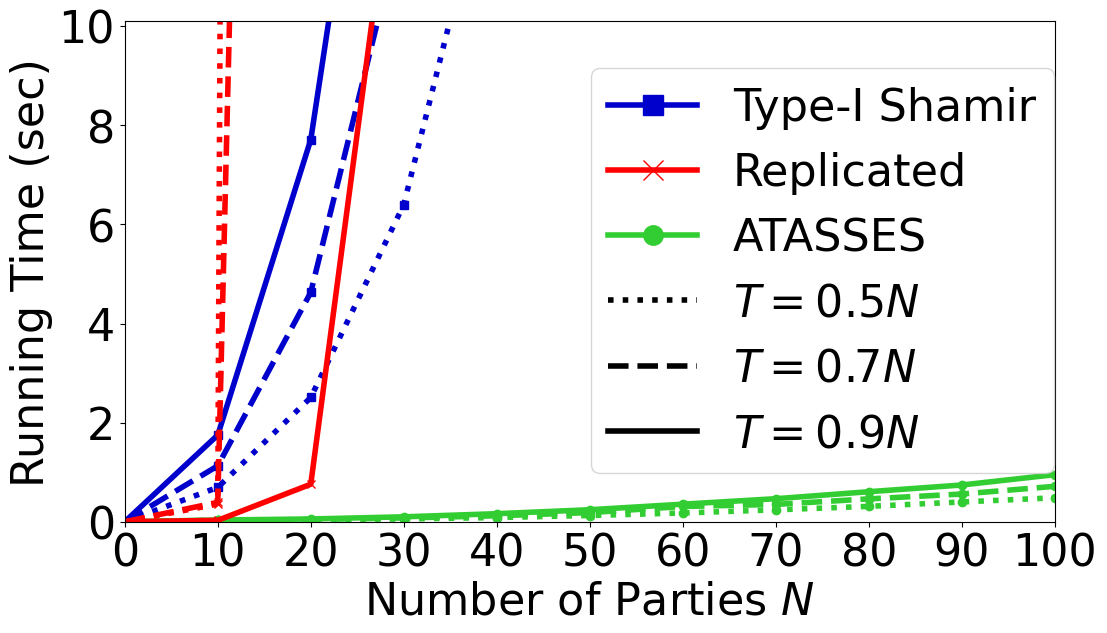}
		\caption{One-round ($K=20M^\prime$)}
		\label{time1k20}
	\end{subfigure}
	\centering
	\begin{subfigure}{0.24\linewidth}
		\centering
		\includegraphics[width=\linewidth]{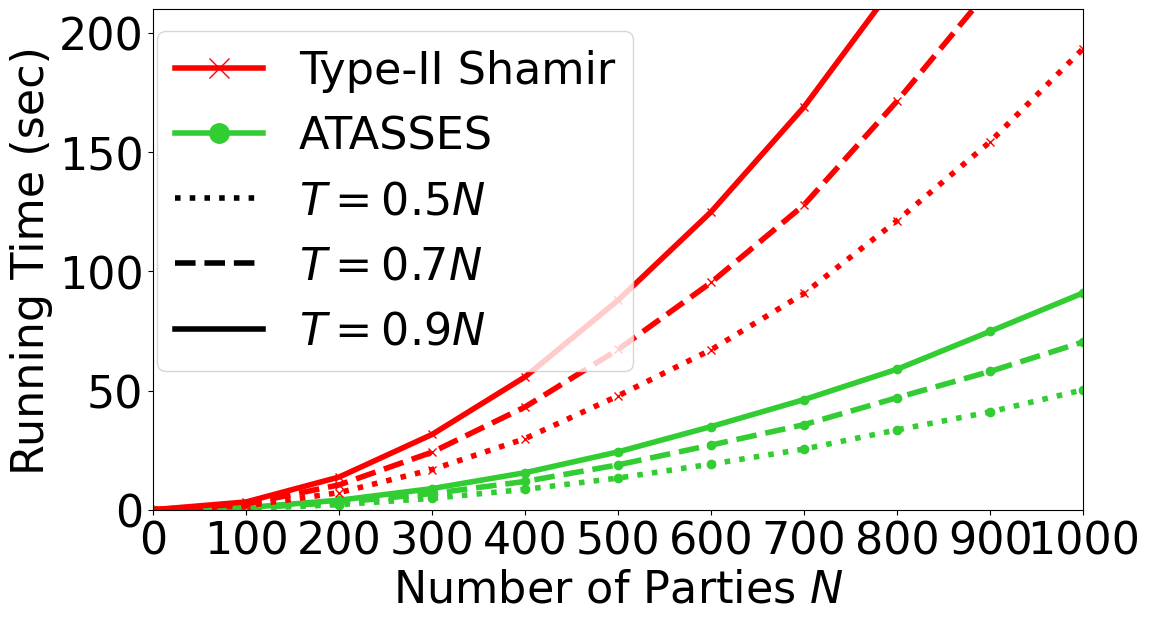}
		\caption{Two-rounds ($K=5M^\prime$)}
		\label{time2k5}
	\end{subfigure}
	\centering
	\begin{subfigure}{0.24\linewidth}
		\centering
		\includegraphics[width=\linewidth]{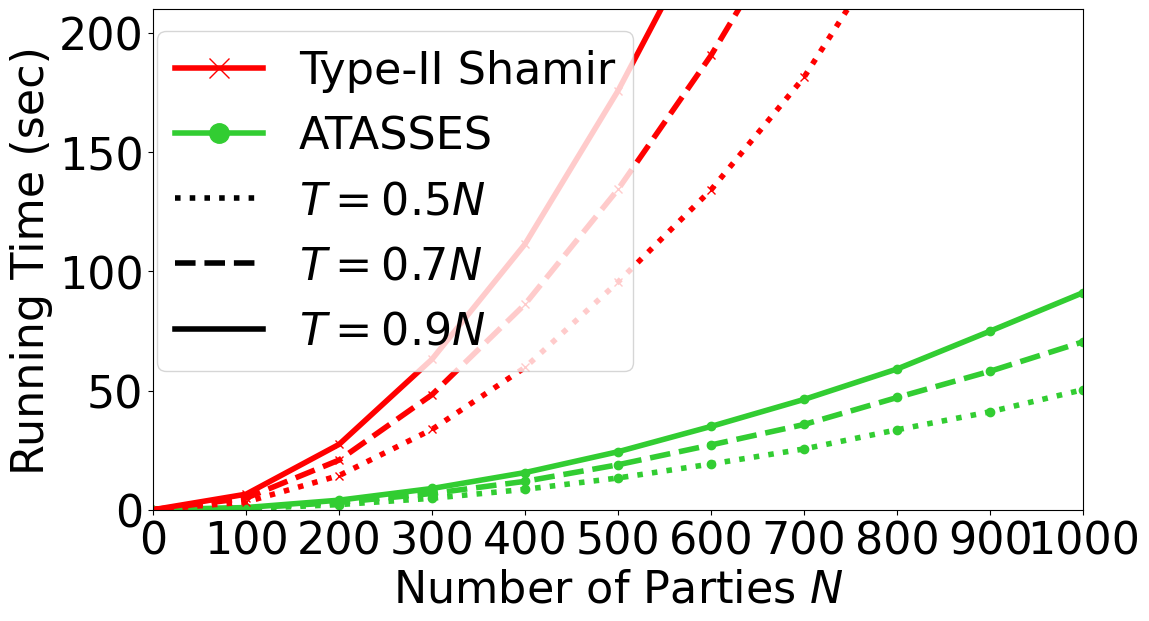}
		\caption{Two-rounds ($K=10M^\prime$)}
		\label{time2k10}
	\end{subfigure}
        \centering
	\begin{subfigure}{0.24\linewidth}
		\centering
		\includegraphics[width=\linewidth]{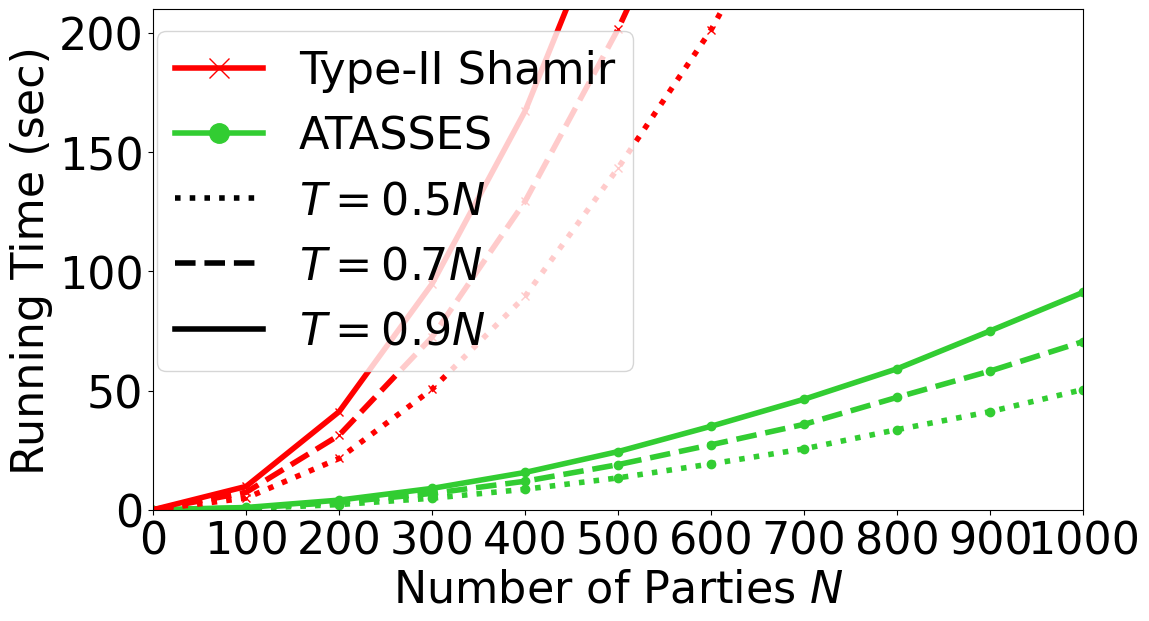}
		\caption{Two-rounds ($K=15M^\prime$)}
		\label{time2k15}
	\end{subfigure}
        \centering
	\begin{subfigure}{0.24\linewidth}
		\centering
		\includegraphics[width=\linewidth]{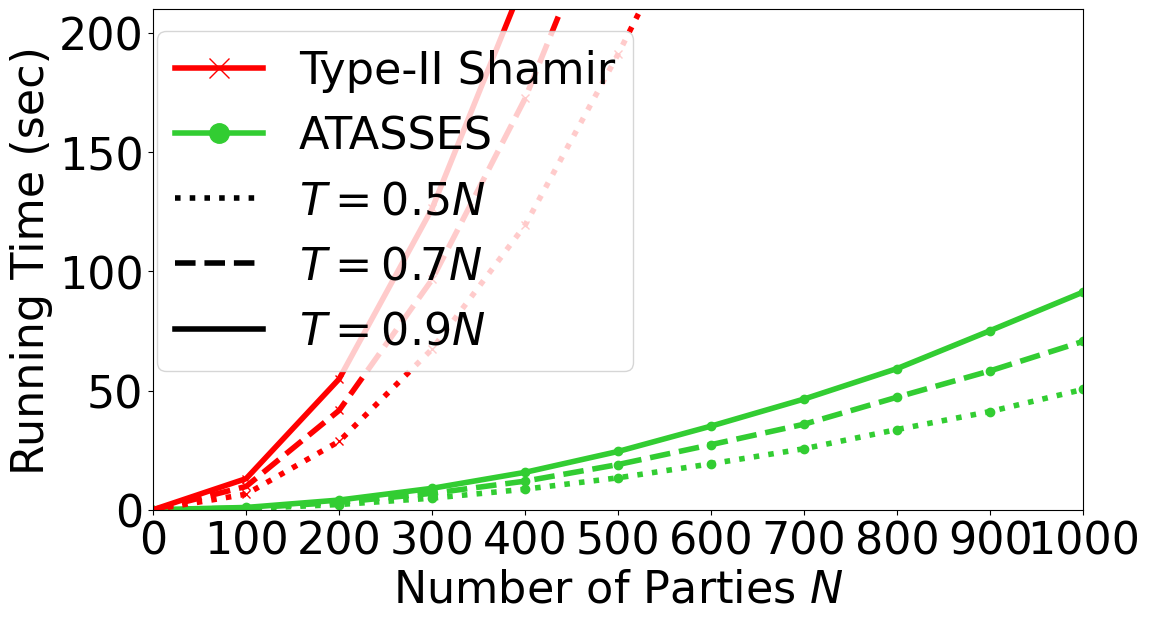}
		\caption{Two-rounds ($K=20M^\prime$)}
		\label{time2k20}
	\end{subfigure}
	\caption{Running time of ATASSES and existing (one-round and two-rounds) ApproxSS with different values of $K$, $N$, and $T$.}
	\label{time}
\end{figure*}
We empirically evaluate ATASSES's efficiency with code in {\color{blue}https://zenodo.org/records/14644655}.

\noindent\textbf{Setup.} We implement the proposed ATASSES as well as three existing ApproxSS schemes for comparison, including two one-round schemes (i.e., replicated ApproxSS as a specific instance of $\{0,1\}$-ApproxSS and Type-\Rmnum{1} Shamir ApproxSS) and a two-round scheme (i.e., Type-\Rmnum{2} Shamir ApproxSS). Particularly, ATASSES relies on the BFV secret-key encryption and is hence implemented on top of Lattigo library~\cite{lattigo}. We note that Type-\Rmnum{3} Shamir ApproxSS has a different setting with others, i.e., it allows the knowledge of participant set in advance. With this setting, Type-\Rmnum{3} Shamir ApproxSS has unfair advantage and is thus excluded from comparison. All experiments in this part are performed in a 14-inch MacBook with Apple M2 Pro CPU.

When implementing these ApproxSS schemes, we adopt the common parameter setting to make a fair comparison. Since ATASSES adopts the default parameter setting $\mathtt{PN12QP109}$ provided by Lattigo library, we set the message space of almost all ApproxSS schemes to be the polynomial ring with degree $4096$ and modulus $65537$. The only exception is Type-\Rmnum{1} Shamir ApproxSS. Since Type-\Rmnum{1} Shamir ApproxSS requires a much larger message space, we use Golang's ``math/big'' package to search the smallest modulus that can meet its requirement. As for the norm bound of noises $B_{sm}$, we set it to be $2^{16}$ for all schemes.\footnote{We note that in practice, this value should be set according to the requirements of applications. For example, the ThFHE may require the value of $B_{sm}$ to be in the polynomial order or even super-polynomial order with respect to some security parameter. In the experiments, we fix the value of $B_{sm}$ for all schemes to make a fair comparison.}

In addition, we adopt multiple values of $T$, $N$, and $K$ to observe their impacts on performance. Specifically, we set $T\in\{0.5N,0.7N,0.9N\}$. The value of $N$ ranges from $10$ to $100$ in existing one-round schemes, while ranges from $100$ to $1000$ in existing two-round schemes. The reason behind such a difference is that those one-round schemes are not efficient enough to support a large number of parties.
As for the value of $K$, we set it to be the multiples of key's size. Let $M^\prime$ denote the size of key used in ATASSES's encryption scheme, we consider $K\in\{5M^\prime,10M^\prime,15M^\prime,20M^\prime\}$.

\noindent\textbf{Results.} We measure the running time of approximate recovery protocols as the performance metric. The running time includes communication time and computation time. The communication time is set to be the size of transferred data divided by the network bandwidth $\mathsf{Band}$. We set $\mathsf{Band}=98$Mbps, which emulates the realistic bandwidth of 4G cellular networks~\cite{5G,bandwidth}. The experiment results are shown in Figure \ref{time}, from which we can come to the following conclusions.

Regarding the growth of running time with $N$, we can observe a slow growth trend of ATASSES. This aligns with our theoretical analysis and demonstrates that ATASSES can be applied to large-scale systems. In contrast, existing ApproxSS schemes have much longer time that grows rapidly with $N$. Particularly, those one-round ApproxSS schemes, despite their lower round-complexity, suffers from extremely-large running time when $N$ exceeds $30$. The two-round ApproxSS scheme performs better, but still worse than ATASSES. More specifically, ATASSES has comparable running time as Type-\Rmnum{2} Shamir ApproxSS when there are a small number of parties, while provides a speedup of $3.83\times$ -- $15.4\times$ when $N=1000$.
Meanwhile, the above observations on comparison between different ApproxSS schemes hold for different values of $T$.

In addition, by observing the same row of sub-figures, the running time of ATASSES hardly changes with the growth of $K$. The reason is that the running time is dominated by the operations whose running time is independent with $K$. In contrast, the running time of existing ApproxSS schemes grows rapidly with different values of $K$. Hence, ATASSES's performance advantage becomes more significant with a larger value of $K$. This demonstrates that ATASSES has the advantage to be applied when the data size is large. Meanwhile, we also observe that ATASSES may lack superiority when $K$ is smaller than the length of BFV's secret key $M^\prime$. Nonetheless, this can be mitigated by choosing an adequate value of $M^\prime$.



\section{Related Work}
\label{sec:related}

In the interest of space, this section reviews related work solely on \textit{arbitrary-threshold} FHE schemes that try to reduce the complexity. Particularly, we divide these schemes into the synchronous and asynchronous settings. 


Most of existing schemes are designed in the asynchronous setting, as it can cover more real-world situations. In this setting, the set of participants can vary at any time, and a participant can NOT learn the set of other participants until receiving some message from them. Asharov et al. proposed a secret key resharing idea to extend a full-threshold ThHE to arbitrary-threshold ThHE~\cite{mpcviaFHE}. Nonetheless, their idea would reveal the secret key of non-participants, resulting in a security vulnerability. Boneh et al. proposed two arbitrary-threshold ThHE schemes based on $\{0,1\}$-linear SS and Shamir SS, respectively~\cite{Threshold}. Although these two schemes avoid the secret key leakage and only require one-round communication for decryption, they both suffer from significant overhead that grows rapidly with the number of parties $N$.
Concurrently to these one-round schemes, Tian et al. design a two-round scheme based on Shamir secret sharing and apply them in the scenario of federated learning~\cite{interactiveTFHE}, which also occurs in~\cite{nuoyafangzhou}. This scheme asks each party to share a secret noise with every parties, resulting in relatively high complexity.

Different from the above work, Mouchet et al. note that the asynchronous setting might be an overkill for many applications and consider to design an efficient scheme in the synchronous setting~\cite{efficient}. In this setting, the participant set can be known to every parties as a prior knowledge.
By utilizing this knowledge, they devise a decryption protocol with $\mathcal{O}(NK)$ computation complexity and $\mathcal{O}(K)$ communication complexity.
Considering that parties in the expected participant set may also crash, 
Helium framework~\cite{Helium} extends \cite{efficient} with a concrete \textit{retry} mechanism.
Specifically, if the expected participant set does not fully match the physical truth,  Helium allows the parties to re-execute the decryption protocol with a modified set of expected participants.
Once the modified set matches the physical truth of participant set, the decryption protocol can output the correct result.
Nonetheless, the successful decryption still requires that the $T$ participants be exactly as expected.
Although this condition can be met when those non-participants become unavailable following \textit{random process} as modeled in Helium's churn model, these works may fail to handle all types of non-participants. For example, a denial-of-service adversary may deliberately switch a party to non-participant once this party is identified as an expected participant.
In this case, the decryption protocol cannot output the correct result even after multiple retries, because the set of expected participants will never match the physical truth.


In summary, existing schemes in the asynchronous setting suffer from low efficiency and existing schemes in the synchronous setting may fail to handle all types of non-participants. In this work, we aim to propose an AThFHE scheme with lower complexity that can successfully decrypt as long as there are at least $T$ participants (could be \textit{any} $T$ parties) at the same time. Meanwhile, we note that some works also try to optimize AThFHE along two other lines. Along one line, recent work~\cite{PELTA} considers \textit{maliciously-secure} ThFHE design. Still, this work assume that there are no non-participants. Our work can help to remove this assumption and advance this work to fully-malicious security.
Along the other line, a series of works~\cite{realworldApp,peter,moduluspolynomial} improve the efficiency by reducing the value of noises. Particularly, they show that the noise smudging technique can also be applied when the noise is in the polynomial order rather than exponential order. Since the ciphertext modulus needs to grow with a larger value of noises, a lower-order noise can reduce the ciphertext modulus from exponential order to polynomial order, which helps to decrease the communication and computation complexity. Notably, our work is compatible with noises of any order and thus can employ these works to reduce the ciphertext modulus.



\section{Conclusion}
\label{sec:conclusion}
We formulate a novel primitive called approximate secret sharing (ApproxSS), and establish the reduction of constructing ThFHE schemes to ApproxSS designs. We develop ATASSES, a Shamir-based ApproxSS scheme in the arbitrary-threshold case with lower complexity. We theoretically prove its security and correctness with guaranteed output delivery, and then empirically demonstrate its substantial efficiency advantages over baselines.
ATASSES helps to induce an efficient arbitrary-threshold ThFHE scheme that outperforms existing schemes for higher efficiency, by which ThFHE can be truly applied in the real world, such as the secure aggregation in federated learning.
Overall, we believe that the proposed primitive can not only improve the efficiency of ThFHE, but also be of independent interest with more applications.

\section{Ethics Considerations and Compliance with Open Science Policy}

We have carefully considered the ethics following the conference guideline.
Our research aims to reduce the complexity of cryptographic primitives by proposing novel techniques.
The involved stakeholders could include the designers of secure multi-party protocols and the participants of computing services. Our work will have positive effects on 1) facilitating the designers with more tools and 2) protecting the privacy of participants.
Meanwhile, both of the research process and our contributions have no negative impacts, including but not limited to breaking the security of computer systems, collecting private information, and violating human rights.

Regarding the open science policy, we fully obey it and have published our code via anonymous link (see Section \ref{sec:experiment}).

\bibliographystyle{plain}

\appendix

\section{Proof of Theorem 1}
\label{appendix-proof-1}

    \begin{thmprime2}{theo:MPHE-correctness}[Correctness of ThFHE, Restated]
    Given a $T$-out-of-$N$ ApproxSS scheme with linearity and $\mathcal{M}_B$-approximate correctness, the ThFHE construction in Figure \ref{fig:MPHE} is a correct $T$-out-of-$N$ scheme with guaranteed output delivery if $B<\Delta/2-B_{\mathsf{CT}}$.  
\end{thmprime2}

\noindent\textit{Proof:} To prove the correctness of ThFHE with guaranteed output delivery, we must show that given a ciphertext $\mathsf{CT}=(\mathbf{c}_0,\mathbf{c}_1)$ as the encryption of $\mathbf{m}$ under the global public key $\mathsf{sk}$, the decryption protocol $\Pi_{\mathsf{Dec}}$ will output $\mathbf{m}$ with overwhelming probability.

In the Phase 1 of decryption protocol, we have $\mathbf{b}_{i}=\mathbf{c}_{1}\cdot\mathsf{skShare}_i+\mathbf{c}_0$. By the linearity of ApproxSS, since $\mathsf{skShare}_i$ is a share of global secret key $\mathsf{sk}$, $\mathbf{b}_i$ is a share of $\mathbf{c}_{1}\cdot\mathsf{sk}+\mathbf{c}_0$. According to the definition of BFV encryption, the ciphertext has the structure $\mathbf{c}_0+\mathbf{c}_1\cdot \mathsf{sk}=\Delta\mathbf{m}+\mathbf{e}_{\mathsf{CT}}$, with $\norm{\mathbf{e}_{\mathsf{CT}}}<B_{\mathsf{CT}}$. Hence, we can learn that $\mathbf{b}_i$ is a share of $\Delta\mathbf{m}+\mathbf{e}_{\mathsf{CT}}$. Let $\mathbf{b}$ denote the $\Delta\mathbf{m}+\mathbf{e}_{\mathsf{CT}}$.

In the Phase 2 of decryption protocol, the approximate recovery protocol is executed to output $\mathbf{b}^\prime$ with only $T$ participants. By the $\mathcal{M}_B$-approximate correctness of ApproxSS, we have $\mathbf{b}^\prime-\mathbf{b}\in\mathcal{M}_B$ and thus $\norm{\mathbf{b}^\prime-\mathbf{b}}<\Delta/2-B_\mathsf{CT}$.
Recall that $\mathbf{b}=\Delta\mathbf{m}+\mathbf{e}_{\mathsf{CT}}$ from the conclusion of Phase 1. Hence, we have 
$\mathbf{b}^\prime=\Delta\mathbf{m}+\mathbf{e}_{\mathsf{CT}}+\mathbf{n}$ for some $\norm{\mathbf{n}}<\Delta/2-B_\mathsf{CT}$.

In the Phase 3 of decryption protocol, the plaintext $\mathbf{m}$ is decoded from $\mathbf{b}^\prime$. By the decryption requirement of BFV, the decryption will succeed to output $\mathbf{m}$ if and only if $\norm{\mathbf{e}_{\mathsf{CT}}+\mathbf{n}}< \Delta/2$. Recall that $\norm{\mathbf{e}_{\mathsf{CT}}}< B_{\mathsf{CT}}$ and $\norm{\mathbf{n}}<\frac{\Delta}{2}-B_\mathsf{CT}$ from the parameter setting and the conclusion of Phase 2. By triangle inequality $\norm{a+b}<\norm{a}+\norm{b}$, we have 
$\norm{\mathbf{e}_{\mathsf{CT}}+\mathbf{n}}<\norm{\mathbf{e}_{\mathsf{CT}}}+\norm{\mathbf{n}}< B_{\mathsf{CT}}+(\Delta/2-B_{\mathsf{CT}})=\Delta/2$. since the condition $\norm{\mathbf{e}_{\mathsf{CT}}+\mathbf{n}}< \Delta/2$ holds, the decryption protocol will successfully output $\mathbf{m}$, which concludes this proof.
$\hfill\blacksquare$

\section{Proof of Theorem 2}
\label{appendix-proof-2}

\begin{thmprime2}{theo:MPHE-security}[Security of ThFHE, Restated]

Given a $T$-out-of-$N$ ApproxSS scheme with vanilla security and $\chi$-approximate security, the ThFHE construction in Figure \ref{fig:MPHE} satisfies simulation-based security if $B_\mathsf{CT}/B_{sm}\in\mathsf{negl}$.
    
\end{thmprime2}

\noindent\textit{Proof:} The proof is conducted via a sequence of hybrid experiments between an adversary $\mathcal{A}$ and a challenger.

\noindent$\mathsf{H}_0$. This is the real-world execution of ThFHE protocols and algorithms. Specifically, the challenger executes the key generation protocol (including the full-threshold FHE's key generation protocol and our $\mathtt{SKShare}$ protocol). Then the challenger encrypts the data of participated parties and send the ciphertexts to the aggregator. Next, the challenger homomorphically evaluates over the ciphertexts from parties and outputs the ciphertext of computation result. Finally, the challenger runs the decryption protocol for parties and aggregator to output the computation result. During this process, the adversary can see the transcripts of every corrupted parties (and in particular, the aggregator) and assign the set of participants in each round.

\noindent$\mathsf{H}_1$. This experiment is the same as $\mathsf{H}_0$, except that the $\mathtt{SKShare}$ in key generation and approximate recovery protocol in decryption are simulated as specified in $\mathsf{Expt}_{\mathcal{A},\mathsf{Ideal}}$. Specifically, the challenger samples a random value $m_\chi$ from $\chi$ and then runs several simulator algorithms to output the transcripts that can be seen by the adversary. By the approximate security of ApproxSS, $\mathsf{H}_1$ and $\mathsf{H}_0$ are indistinguishable.

\noindent$\mathsf{H}_2$. This experiment is the same as $\mathsf{H}_1$, except that the random value $m_\chi$ is simulated by sampling $\mathbf{e}^\prime\sample\{\mathbf{e}\mid\norm{\mathbf{e}}<B_{sm}\}$ and computing $m_\chi=\Delta\cdot\mathbf{m}-\mathbf{c}_0+\mathbf{e}^\prime$, rather than sampling as $\mathbf{b}+\mathbf{x}$ with $\mathbf{x}\sample\chi$. Recall that we set $\chi=\mathsf{Uniform}(\mathcal{M}_{B_{sm}}=\{\mathbf{e}\mid\norm{\mathbf{e}}<B_{sm}\})$. To demonstrate that $\mathsf{H}_1$ and $\mathsf{H}_2$ are indistinguishable, it suffices to show that $\mathbf{b}+\mathbf{x}$ is indistinguishable from $\Delta\cdot\mathbf{m}-\mathbf{c}_0+\mathbf{e}^\prime$ with $\mathbf{e}^\prime\sample\{\mathbf{e}\mid\norm{\mathbf{e}}<B_{sm}\}$. Note that $\mathbf{b}=\mathbf{c}_1\cdot \mathsf{sk}$ and thus $\mathbf{b}+\mathbf{e}=\mathbf{c}_1\cdot \mathsf{sk}+\mathbf{e}=\Delta\cdot\mathbf{m}-\mathbf{c}_0+\mathbf{e}_{\mathsf{CT}}+\mathbf{e}$. When $B_{\mathsf{CT}}/B_{sm}\in\mathsf{negl}$, we have that $\mathbf{e}_{\mathsf{CT}}+\mathbf{e}$ is indistinguishable with $\mathbf{e}^\prime$ by the smudging lemma~\cite{mpcviaFHE}. Hence, $\chi$ is indistinguishable from $\Delta\cdot\mathbf{m}-\mathbf{c}_0+\mathbf{e}^\prime$ with $\mathbf{e}^\prime\sample\{\mathbf{e}\mid\norm{\mathbf{e}}<B_{sm}\}$ and thus $\mathsf{H}_1$ and $\mathsf{H}_2$ are indistinguishable.

\noindent$\mathsf{H}_3$. This experiment is the same as $\mathsf{H}_2$, except that the challenger does not generate the parties' ciphertext by encrypting their data. Instead, the challenger generates them by encrypting $0$. By the CPA security of BFV encryption, $\mathsf{H}_3$ is indistinguishable with $\mathsf{H}_2$.

We note that $\mathsf{H}_3$ is the ideal-world experiment. The reason is that $\mathsf{H}_3$ simulates the view of adversary without using the private data of parties. Instead, the only used data is public parameters and the final output $\mathbf{m}$.

By the above arguments, the real-world experiment $\mathsf{H}_0$ is indistinguishable from the ideal-world experiment $\mathsf{H}_3$.
Hence, ThFHE satisfies the simulation-based security.
$\hfill\blacksquare$

\section{Proof of Theorem 3}

\label{appendix:proof-atasses}
\begin{thmprime2}{theo:ATASSES}[Properties of ATASSES, Restated]

ATASSES satisfies $\chi$-approximate correctness and $\mathcal{M}_B$-approximate security under RLWE-hardness assumption for $\chi=\mathsf{Uniform}(\{n\mid \norm{n}\leq B_{sm}\})$ and $\mathcal{M}_B=\{n\mid \norm{n}\leq T\cdot B_{sm}\}$.
    
\end{thmprime2}

\noindent\textit{Proof:} This proof consists of two parts, which prove the approximate correctness and the approximate security of ATASSES, respectively.

\noindent\textbf{Approximate Correctness:} We first prove the approximate correctness of ATASSES. By Definition \ref{def:approxcorrect}, the approximate correctness says that the output of $\Pi_{\mathsf{ApproxRec}}$, i.e., $m^\prime$, should satisfy $m^\prime=m+n$ for some $n\in\mathcal{M}_B$. Since ATASSES sets $\mathcal{M}_B=\{n\mid \norm{n}\leq T\cdot B_{sm}\}$, we just need to prove $m^\prime=m+\sum_{i\in\mathcal{T}}n_i$ with $\norm{n_i}\leq B_{sm}, \forall i$. Since $m^\prime$ is obtained by decrypting $\mathsf{CTall}$ with key $\mathsf{dk}$, it suffices to prove that $\mathsf{CTall}$ is truly the encryption of $m+\sum_{i\in\mathcal{T}}n_i$ using $\mathsf{dk}$. Next, we prove this conclusion by proving two intermediate conclusions: 1) the $\mathsf{CTall}$ can be decrypted to $m+\sum_{i\in\mathcal{T}}n_i$ using the key $\sum_{i\in\mathcal{T}} L_i^{(\mathcal{T}_i)} \cdot \mathsf{ek}_{i,1}+\mathsf{ek}_{i,2}$ and 2) $\mathsf{dk}=\sum_{i\in\mathcal{T}} L_i^{(\mathcal{T}_i)} \cdot \mathsf{ek}_{i,1}+\mathsf{ek}_{i,2}$.

\textit{Proof for intermediate conclusion 1.} By the algorithm $\mathtt{PartyR1}$, the ciphertexts are obtained as
\begin{align}
    \mathsf{CTs}_i&=(\mathbf{a}\cdot \mathsf{ek}_{i,1}+\mathbf{e}_{i,1}+\Delta^\prime \cdot s_i,-\mathbf{a});\label{eq:idea-CT-1-appendix}\\
    \mathsf{CTn}_i&=(\mathbf{a}\cdot \mathsf{ek}_{i,2}+\mathbf{e}_{i,2}+\Delta^\prime \cdot n_i,-\mathbf{a});\label{eq:idea-CT-2-appendix}\\
    \mathsf{CT}_{all}&=(\sum_{i\in\mathcal{T}} (L_i\cdot\mathsf{CTs}_i[0]+\mathsf{CTn}_i[0]),-\mathbf{a})\label{eq:idea-CT-3}.
\end{align}
By substituting Eq.\eqref{eq:idea-CT-1-appendix} and Eq.\eqref{eq:idea-CT-2-appendix} to Eq.\eqref{eq:idea-CT-3}, decrypting $\mathsf{CT}_{all}$ using $\sum_{i\in\mathcal{T}} L_i^{(\mathcal{T}_i)} \cdot \mathsf{ek}_{i,1}+\mathsf{ek}_{i,2}$ leads to 
\begin{align}
    &\mathtt{BFV.Dec}(\sum_{i\in\mathcal{T}} L_i^{(\mathcal{T}_i)} \cdot \mathsf{ek}_{i,1}+\mathsf{ek}_{i,2},\mathsf{CT}_{all})\\
    =&\sum_{i\in\mathcal{T}} (L_i\cdot\mathsf{CTs}_i[0]+\mathsf{CTn}_i[0])-\mathbf{a}\cdot(\sum_{i\in\mathcal{T}} L_i^{(\mathcal{T}_i)} \cdot \mathsf{ek}_{i,1}+\mathsf{ek}_{i,2})\\
    =&\sum_{i\in\mathcal{T}} [\Delta^\prime\cdot(L_is_i+n_i)+(L_i\mathbf{e}_{i,1}+\mathbf{e}_{i,2})]\label{eq:idea-1}\\
    =&\Delta^\prime\cdot(m+\sum_{i\in\mathcal{T}} n_i)+\sum_{i\in\mathcal{T}}(L_i\mathbf{e}_{i,1}+\mathbf{e}_{i,2}).
\end{align}
Recall that we set $\Delta^\prime/2>N\cdot P^\prime\cdot B^\prime$. Hence, the norm of total error $\sum_{i\in\mathcal{T}}(L_i\mathbf{e}_{i,1}+\mathbf{e}_{i,2})$ must be bounded by $N\cdot P^\prime\cdot B^\prime$ as well as $\Delta^\prime/2$. This aligns with the requirement of successful BFV decryption. Hence, decrypting $\mathsf{CT}_{all}$ using $\sum_{i\in\mathcal{T}} L_i^{(\mathcal{T}_i)} \cdot \mathsf{ek}_{i,1}+\mathsf{ek}_{i,2}$ leads to $m+\sum_{i\in\mathcal{T}} n_i$. The first intermediate conclusion is proved.

\textit{Proof for intermediate conclusion 2.} Note that $\mathsf{dk}$ is generated by running the recovery algorithm of Shamir secret sharing over shares $\{\mathsf{dkShare}_j\}_{j\in\mathcal{T}_2}$. Each decryption key share is generated by $\mathsf{dkShare}_j=\sum_{i\in\mathcal{T}} L_i^{(\mathcal{T})}\cdot\mathsf{ekShare}_{i,j,1}+\mathsf{ekShare}_{i,j,2}$. By the linearity of Shamir secret sharing, $\mathsf{dk}$ is equal to $\sum_{i\in\mathcal{T}} L_i^{(\mathcal{T})}\cdot\mathsf{ek}_{i,1}+\mathsf{ek}_{i,2}$. The second intermediate conclusion is proved.

Based on the above two intermediate conclusions, $\mathsf{CTall}$ is truly the encryption of $m+\sum_{i\in\mathcal{T}}n_i$ using $\mathsf{dk}$. Hence, $m^\prime=m+n$ for some $n\in\mathcal{M}_B=\{n\mid \norm{n}\leq T\cdot B_{sm}\}$. The approximate correctness of ATASSES holds.

\noindent\textbf{Approximate Security:} Below, we first prove a property of RLWE secret-key encryption and then use it to prove the semi-honest security of ATASSES. Our proof relies on the following lemma from the full version of~\cite{acorn}.



\begin{lemma}[Lemma 4, Appendix A in~\cite{acorn}]
\label{lemma:4}
    For any $\sigma_1>0$, for any $m,K,Q,P,\Delta,l\geq 1$, let $k=l/K$ and $x_1,\ldots,x_m\in\mathbb{Z}_P^l\equiv \mathcal{R}_P^{k}$. Assume $\mathsf{RLWE}_{K,q,\sigma}$ is hard for $\sigma=1/\sqrt{2}\sigma_1$. Then, the following two distributions $D_0$ and $D_1$ are indistinguishable.
\end{lemma}

$$
D_0=\left\{
\begin{aligned}
             (A,&\,As_1+(e_1+f_1)+\Delta x_1,\ldots,  \\
             &As_{m-1}+(e_{m-1}+f_{m-1})+\Delta x_{m-1},\\
             &-A\sum_{i=1}^{m-1} s_i+(e_m+f_m)+\Delta x_m )\mod{Q}:\\
             A\gets&\mathcal{R}_Q^k,s_1,\ldots,s_{m-1}\sample \chi_s, e_i, f_i\sample D_{\sigma_1}^k, \forall i
\end{aligned}
 \right\}
     $$   

$$
D_1=\left\{
\begin{aligned}
             (A,&\,u_1,\ldots,u_{m-1},  \\
             &-\sum_{i=1}^{m-1} u_i+\sum_{i=1}^{m}(e_i+f_i)+\Delta \sum_{i=1}^{m}x_i) \mod{Q}:\\
             A\gets&\mathcal{R}_Q^k,u_1,\ldots,u_{m-1}\sample \mathcal{R}_Q^k, e_i, f_i\sample D_{\sigma_1}^k,\forall i
\end{aligned}
 \right\}
     $$   

Intuitively, this lemma demonstrates that the joint distribution of the ciphertexts of $x_i$s ($D_0$) is indistinguishable from that of random values ($D_1$), conditioned on the sum of these ciphertexts is the same as the sum of random values. To apply this lemma to our proof, we slightly modify it by considering the indistinguishability between two new distributions $D_0^\prime$ and $D_1^\prime$. The differences between $D_0$ and $D_0^\prime$, as well as $D_1$ and $D_1^\prime$, are highlighted in red color. Basically, $D_0^\prime$ (resp. $D_1^\prime$) contains an extra random value $As$ and this $As$ is added to the last ciphertext in $D_0^\prime$ (resp. the last random value in $D_1^\prime$). By Lemma \ref{lemma:4}, $D^\prime_0$ and $D^\prime_1$ are also indistinguishable.

$$
D_0^\prime=\left\{
\begin{aligned}
             (A,&\,{\color{red}As},As_1+(e_1+f_1)+\Delta x_1,\ldots,  \\
             &As_{m-1}+(e_{m-1}+f_{m-1})+\Delta x_{m-1},\\
             &{\color{red}As-}A\sum_{i=1}^{m-1} s_i+(e_m+f_m)+\Delta x_m )\mod{Q}:\\
             A\sample&\mathcal{R}_Q^k,s,s_1,\ldots,s_{m-1}\sample \chi_s, e_i, f_i\sample D_{\sigma_1}^k, \forall i
\end{aligned}
 \right\}
     $$   

$$
D_1^\prime=\left\{
\begin{aligned}
             (A,&\,{\color{red}As},u_1,\ldots,u_{m-1},  \\
             &{\color{red}As-}\sum_{i=1}^{m-1} u_i+\sum_{i=1}^{m}(e_i+f_i)+\Delta \sum_{i=1}^{m}x_i) \mod{Q}:\\
             A\sample&\mathcal{R}_Q^k,{\color{red}s\sample\chi_s},u_1,\ldots,u_{m-1}\sample \mathcal{R}_Q^k, e_i, f_i\sample D_{\sigma_1}^k,\forall i
\end{aligned}
 \right\}
     $$   

Intuitively, the indistinguishability between $D_0^\prime$ and $D_1^\prime$ states that the joint distribution of the ciphertexts of $x_1,\ldots,x_m$s encrypted by different encryption keys $s_1,\ldots,s_{m-1},s-\sum_{i\in[m-1]} s_i$ is indistinguishable from that of random values, conditioned on 1) the sum of these ciphertexts is the same as the sum of random values and 2) the sum of encryption keys $s$ is publicly known. In other words, the sum of encryption keys is safe to disclose without revealing the encryption keys and the encrypted messages.

Next, we prove the approximate security of ApproxSS based on this conclusion. Note that ATASSES consists of two rounds. By Definition \ref{def:approx-security}, to prove the approximate security of ATASSES, we need to prove the existence of simulator algorithms $\{S_0,S_1,S_2\}$, so that the real-world experiment $\mathsf{Expt}_{\mathcal{A},\mathsf{Real}}$ and ideal-world experiment $\mathsf{Expt}_{\mathcal{A},\mathsf{Ideal}}$ are indistinguishable. Next, the proof proceeds by first describing the simulator algorithms $\{S_0,S_1,S_2\}$ and then constructing successive hybrid experiments between $\mathsf{Expt}_{\mathcal{A},\mathsf{Real}}$ and $\mathsf{Expt}_{\mathcal{A},\mathsf{Ideal}}$.  

Before describing the simulator algorithms, we give some notations.
Let $\mathcal{A}_r$ and $\mathcal{H}_r$ denote the set of corrupted participants and the set of honest participants in round $r$, respectively. Hence, the set $\mathcal{T}_r$ of participants in round $r$ satisfies $\mathcal{T}_r=\mathcal{A}_r\cup\mathcal{H}_r$. Note that there exists at least one honest participant by our system model. We use $h_r$ to denote this participant in round $r$. In addition, we use the same notation as Section 2.3 in~\cite{acorn} and rewrite an RLWE sample from $\mathbf{a}\cdot\mathsf{sk}+\mathbf{e}$ to $A\cdot sk+e$, where $A,sk,e$ are matrices and vectors generated from the coefficient embedding of polynomials $\mathbf{a},\mathsf{sk},\mathbf{e}$. Such a rewritting helps to simplify the description with long message. Recall that in ATASSES, the message $m_i$ can be too long to be encrypted into one ciphertext. Hence, party $i$ needs to break down $m_i$ into $C^\prime$ sub-vectors and encrypt each sub-vector $m_{i,k}$ using $\mathbf{a}_k$ and $\mathbf{e}_k$, respectively. By using the matrix-vector notation, we can rewrite the encryption of long message $m_i$ as $A\cdot s+e+\Delta m_i$, where $A$ and $e$ are the concatenation of $A_1,\ldots,A_{C^\prime}$ and $e_1,\ldots,e_{C^\prime}$, respectively.

\noindent\textit{Simulator algorithm $S_0(N,T,\mathcal{M})$:} This algorithm outputs $\{s_i\}_{i\in\mathcal{A}}$ by sampling $s_i$s from $\mathsf{Uniform}(\mathcal{M})$.

\noindent\textit{Simulator algorithm $S_1(m_{\chi},\mathcal{T}_1)$:} Recall that algorithm $\mathtt{PartyR1}$ of ATASSES asks each party $i\in\mathcal{T}_1$ to generate two things: 1) the shares of encryption keys $\mathsf{ekShare}_{j,i,1}$ and $\mathsf{ekShare}_{j,i,2}$ and 2) the ciphertexts $\mathsf{CTs}_{i,k}$ and $\mathsf{CTn}_{i,k}$. The $S_1$ simulates these outputs as follows:
\begin{itemize}
    \item If $i\in\mathcal{A}_1$, $S_1$ follows algorithm $\mathtt{PartyR1}()$ in the real world to generate those outputs.
    \item If $i\in\mathcal{H}_1\setminus\{h_1\}$, $S_1$ samples its encryption key shares $\mathsf{ekShare}_{j,i,1}$ and $\mathsf{ekShare}_{j,i,2}$ for all $j\in[N]$ from the uniform distribution over the ciphertext space. As for the ciphertexts, $S_1$ uses the same $A\sample\mathsf{CRS}$ as $\mathsf{CTs}_{i}[1]$ and $\mathsf{CTn}_{i}[1]$ for all $i$, but samples uniformly-random elements from ciphertext space as $\mathsf{CTs}_{i}[0]$ and $\mathsf{CTn}_{i}[0]$.
    \item If $i=h_1$, $S_1$ samples its encryption key shares $\mathsf{ekShare}_{j,h_1,1}$ and $\mathsf{ekShare}_{j,h_1,2}$ for all $j\in[N]$ from the uniform distribution over the ciphertext space. As for the ciphertexts, $S_1$ uses the same $A\sample\mathsf{CRS}$ as $\mathsf{CTs}_{h_1}[1]$ and $\mathsf{CTn}_{h_1}[1]$ and samples uniformly-random elements from ciphertext space as $\mathsf{CTs}_{h_1}[0]$. As for $\mathsf{CTn}_{h_1}[0]$, $S_1$ generates it as follows:
    \begin{equation}
    \begin{split}
       &\mathsf{CTn}_{h_1}[0]=A\cdot s+\sum_{i\in\mathcal{T}_1}L_i^{(\mathcal{T}_1)} e_{i,1}+\sum_{i\in\mathcal{H}_1} e_{i,2}\\
       &+\Delta^\prime\cdot (m_{\chi}-\sum_{i\in\mathcal{A}_1} L_i^{(\mathcal{T}_1)} s_{i}+\sum_{i\in\mathcal{H}_1\setminus\{h_1\}} n_i )
        \\
        &-\sum_{i\in\mathcal{T}_1}L_i^{(\mathcal{T}_1)} \cdot\mathsf{CTs}_{i}[0]-\sum_{i\in\mathcal{H}_1\setminus\{h_1\}} \mathsf{CTn}_{i}[0]
        \end{split}
    \end{equation}
    where $s$ is sampled from the key distribution, $e_{i,1}$, $e_{i,2}$ are the random errors sampled from the distribution as $e_i+f_i$, and $n_i$s are the random noises sampled from $\{n\mid\norm{n}<B_{sm}\}$.
\end{itemize}

\noindent\textit{Simulator algorithm $S_2(m_\chi,\mathcal{T}_2)$:} Recall that algorithm $\mathtt{PartyR2}$ of ATASSES asks each party $j\in\mathcal{T}_2$ to generate a decryption share $\mathsf{dkShare}_j$. The $S_2$ simulates $\{\mathsf{dkShare}_j\}_{j\in\mathcal{T}_2}$ as follows:
\begin{itemize}
    \item If $j\in\mathcal{A}_2$, $S_2$ follows algorithm $\mathtt{PartyR2}()$ in the real world to output $\mathsf{dkShare}_j$.
    \item If $j\in\mathcal{H}_2\setminus\{h_2\}$, $S_2$ samples $\mathsf{dkShare}_j$ from the uniform distribution over the ciphertext space.
    \item If $j=h_2$, $S_2$ generates $\mathsf{dkShare}_{h_2}$ as 
    \begin{equation}
        (L_{h_2}^{(\mathcal{T}_2)})^{-1}\cdot (s+\sum_{i\in\mathcal{A}_1} \mathsf{ek}_{i,2}- \sum_{j\in\mathcal{T}_2\setminus\{h_2\}} L_j^{(\mathcal{T}_2)} \mathsf{dkShare}_j).
    \end{equation}
\end{itemize}

Based on the above simulator algorithms, we construct the following hybrid experiments by gradually replacing the real-world algorithms by the simulator algorithms. 

\noindent$\mathsf{H}_1$: this is same as $\mathsf{Expt}_{\mathcal{A},\mathsf{Real}}$, except that the generation of ciphertexts in algorithm $\mathsf{PartyR1}$ and algorithm $\mathsf{PartyR2}$ are replaced by simulator algorithms $S_1$ and $S_2$. Notably, the encryption key shares in algorithm $\mathsf{PartyR1}$ remain the same as in $\mathsf{Expt}_{\mathcal{A},\mathsf{Real}}$. 

The difference between $\mathsf{Expt}_{\mathcal{A},\mathsf{Real}}$ and $\mathsf{H}_1$ can be summarized in the form of $D_0^\prime$ and $D_1^\prime$. 
\begin{itemize}
    \item For $\mathsf{H}_1$: In $S_1$, we remark that $\{L_i^{\mathcal{T}_1}\cdot\mathsf{CTs}_{i,k}[0]\}_{i\in\mathcal{T}_1}\cup\{\mathsf{CTn}_{i,k}[0]\}_{i\in\mathcal{H}_1\setminus\{h_1\}}$ plays the role of the former $m-1$ random values $u_1,\ldots,u_{m-1}$ in distribution $D_1^\prime$. The $\mathsf{CTn}_{h_1}[0]$ corresponds to the last random value in $D_1^\prime$. Particularly, $(m_{\chi}-\sum_{i\in\mathcal{A}_1} L_i^{(\mathcal{T}_1)} s_{i}+\sum_{i\in\mathcal{H}_1\setminus\{h_1\}} n_i )$ corresponds to $\sum_{i\in[m]} x_i$, and $\sum_{i\in\mathcal{T}_1}L_i^{(\mathcal{T}_1)} e_{i,1}+\sum_{i\in\mathcal{H}_1} e_{i,2}$ corresponds to $\sum_{i\in[m]} (e_i+f_i)$. In $S_2$, $\sum_{j\in\mathcal{T}_2} L_j^{(\mathcal{T}_2)}\cdot\mathsf{dkShare}_j-\sum_{i\in\mathcal{A}_1}\mathsf{ek}_{i,2}$ corresponds to $s$, and thus, $A\cdot(\sum_{j\in\mathcal{T}_2} L_j^{(\mathcal{T}_2)}\cdot\mathsf{dkShare}_j-\sum_{i\in\mathcal{A}_1}\mathsf{ek}_{i,2})$ corresponds to $As$.
    \item For $\mathsf{Expt}_{\mathcal{A},\mathsf{Real}}$, we remark that $\{L_i^{\mathcal{T}_1}\cdot\mathsf{CTs}_{i,k}[0]\}_{i\in\mathcal{T}_1}\cup\{\mathsf{CTn}_{i,k}[0]\}_{i\in\mathcal{H}_1\setminus\{h_1\}}$ plays the role of the former $m-1$ ciphertexts. By the proof of approximate correctness, we have $\mathsf{dk}=\sum_{i\in\mathcal{T}_1} L_i^{(\mathcal{T}_1)}\cdot\mathsf{ek}_{i,1}+\mathsf{ek}_{i,2}$ and 
    $\mathsf{dk}=\sum_{j\in\mathcal{T}_2} L_j^{(\mathcal{T}_2)}\cdot\mathsf{dkShare}_j$. By combining these two equations, we have $\mathsf{ek}_{h_2,2}=\sum_{j\in\mathcal{T}_2} L_j^{(\mathcal{T}_2)}\cdot\mathsf{dkShare}_j-\sum_{i\in\mathcal{T}_1} L_i^{(\mathcal{T}_1)}\cdot\mathsf{ek}_{i,1}-\sum_{i\in\mathcal{T}_1\setminus\{h_1\}} \mathsf{ek}_{i,2}$, which corresponds to $s-\sum_{i\in\mathcal{T}_1} L_i^{(\mathcal{T}_1)}\cdot\mathsf{ek}_{i,1}-\sum_{i\in\mathcal{H}_1\setminus\{h_1\}} \mathsf{ek}_{i,2}$ in $D_0^\prime$. Hence, $\sum_{j\in\mathcal{T}_2} L_j^{(\mathcal{T}_2)}\cdot\mathsf{dkShare}_j-\sum_{i\in\mathcal{A}_1}\mathsf{ek}_{i,2}$ corresponds to $s$, and thus, $A\cdot(\sum_{j\in\mathcal{T}_2} L_j^{(\mathcal{T}_2)}\cdot\mathsf{dkShare}_j-\sum_{i\in\mathcal{A}_1}\mathsf{ek}_{i,2})$ corresponds to $As$.
\end{itemize}

By the above arguments on differences, to distinguish between $\mathsf{Expt}_{\mathcal{A},\mathsf{Real}}$ and $\mathsf{H}_1$, the adversary is in fact to distinguish two distributions in the form of $D_0^\prime$ and $D_1^\prime$. Nonetheless, $D_0^\prime$ and $D_1^\prime$ are indistinguishable under RLWE-hardness assumption.
Hence, $\mathsf{Expt}_{\mathcal{A},\mathsf{Real}}$ and $\mathsf{H}_1$ are indistinguishable.

\noindent$\mathsf{H}_2$: this is same as $\mathsf{H}_1$, except that the encryption key shares is generated by simulator algorithm $S_0$ instead of $\mathtt{PartyR1}$.

The only difference between $\mathsf{H}_2$ and $\mathsf{H}_1$ is the encryption key shares of parties corrupted by $\mathcal{A}$. We note that by the security of Shamir secret sharing, the $T$-out-of-$N$ shares from $T-1$ parties are indistinguishable from uniformly-random values. Since $\mathcal{A}$ can corrupt at most $T-1$ parties by our threat model, $\mathsf{H}_1$ and $\mathsf{H}_2$ are indistinguishable.

\noindent$\mathsf{H}_3$: this is same as $\mathsf{H}_2$, except that the algorithm $\mathsf{Share}$ is replaced by simulator algorithm $S_0$. We note that $\mathsf{H}_3$ is equivalent to $\mathsf{Expt}_{\mathcal{A},\mathsf{Ideal}}$.

The only difference between $\mathsf{H}_3$ and $\mathsf{H}_2$ is the shares of parties corrupted by $\mathcal{A}$. We note that by the security of Shamir secret sharing, the $T$-out-of-$N$ shares from $T-1$ parties are indistinguishable from uniformly-random values. Since $\mathcal{A}$ can corrupt at most $T-1$ parties by our threat model, $\mathsf{H}_1$ and $\mathsf{H}_2$ are indistinguishable.

Based on the above arguments, $\mathsf{Expt}_{\mathcal{A},\mathsf{Real}}$ and $\mathsf{Expt}_{\mathcal{A},\mathsf{Ideal}}$ are indistinguishable and thus ATASSES satisfies approximate security.
$\hfill\blacksquare$

\section{Complexity Analysis}
\label{appendix-proof-4}
Below we analyze the computation and communication complexity of four ApproxSS schemes, namely $\{0,1\}$-ApproxSS, Type-\Rmnum{1} Shamir ApproxSS, Type-\Rmnum{2} Shamir ApproxSS, and ATASSES. In the analysis, we set $T$ as the order of $\mathcal{O}(N)$, as $T$ is usually set to be the proportion of $N$, say $0.5N$ or $0.9N$. The computation complexity is set to be the summation of computation complexity of the aggregator and a single party. The reason is that the parties execute their algorithms in parallel. Hence, the computation time should be counted as the computation time of the slowest party, rather than the summation of all parties. The communication complexity is set to be the size of transferred message between a pair of parties (or between a party and an aggregator), rather than the size of transferred message between a party and every parties. The reason is that we assume a party has the communication channel with every parties (and the aggregator). Hence, the messages to $N$ parties can be transferred in parallel, rather than to each party one by one.

\noindent\textbf{$\{0,1\}$-ApproxSS}: Existing work~\cite{Threshold} shows that the size of each share of $\{0,1\}$-ApproxSS is $\mathcal{O}(N^{4.2})$ on average. When the data has the size of $K$, the share size is $\mathcal{O}(N^{4.2}\cdot K)$. The computation workload of each party is to sample a noise whose size is $\mathcal{O}(N^{4.2}\cdot K)$ and to add the share and the noise together. Hence, the computation complexity of each party is $\mathcal{O}(N^{4.2}\cdot K)$. The computation workload of the aggregator is to recover the approximate message from $T$ noisy shares. Hence, the computation complexity of the party is $T\cdot\mathcal{O}(N^{4.2}\cdot K)=\mathcal{O}(N^{5.2}\cdot K)$. Therefore, the overall computation complexity is $\mathcal{O}(N^{5.2}\cdot K)$. As for the communication complexity, each party needs to send the noisy share to the aggregator, with the complexity being $\mathcal{O}(N^{4.2}\cdot K)$. Hence, the communication complexity is $\mathcal{O}(N^{4.2}\cdot K)$.

\noindent\textbf{Type-\Rmnum{1} Shamir ApproxSS}: When applying Type-\Rmnum{1} Shamir ApproxSS, the message space has a modulus $\mathcal{O}(N\cdot(N!)^3)$. Hence, the size of each element is $\mathcal{O}(\log (N\cdot(N!)^3))$. By Stirling's formula, the element's size can be approximated to $\mathcal{O}(N\log N)$, which is $\mathcal{O}(N)$ times the element's size of other ApproxSS schemes. Hence, the size of each share is regarded as $\mathcal{O}(NK)$. The computation workload is dominated by the aggregator, who needs to compute Lagrange coefficients with complexity $\mathcal{O}(N^2)$ and compute the linear combination of $T$ shares with complexity $T\cdot\mathcal{O}(NK)=\mathcal{O}(N^2K)$. Hence, the computation complexity is $\mathcal{O}(N^2K)$. As for the communication complexity, each party needs to send the noisy share to the aggregator, with the complexity being $\mathcal{O}(NK)$. Hence, the communication complexity is $\mathcal{O}(NK)$.

\noindent\textbf{Type-\Rmnum{2} Shamir ApproxSS}: Type-\Rmnum{2} Shamir ApproxSS consists of two rounds. In the first round, each party generates the shares of a length-$K$ noise and sends a share to each party with computation complexity $\mathcal{O}(N^2K)$ and communication complexity $\mathcal{O}(NK)$. In the second round, each party adds $T$ noises and the share together and sends the noisy share to the aggregator, resulting in computation complexity $\mathcal{O}(N^2K)$ and communication complexity $\mathcal{O}(NK)$. For the aggregator, to recover the approximate message, it needs to compute Lagrange coefficients with complexity $\mathcal{O}(N^2)$ and compute the linear combination of $T$ shares with complexity $T\cdot\mathcal{O}(NK)=\mathcal{O}(N^2K)$. Hence, the computation complexity is $\mathcal{O}(N^2K)$. In total, the computation complexity is $\mathcal{O}(N^2K)$ and the communication complexity is $\mathcal{O}(NK)$.

\noindent\textbf{Type-\Rmnum{3} Shamir ApproxSS}: Type-\Rmnum{3} Shamir ApproxSS requires only one round. Each party computes a $T$-out-of-$T$ secret key share with computation complexity $\mathcal{O}(N)$ and then generates a decryption share using this secret key share with computation complexity $\mathcal{O}(K)$. Hence, the total computation complexity is $\mathcal{O}(N+K)$. Then the party needs to send this decryption share to the aggregator with communication complexity $\mathcal{O}(K)$. Next, the aggregator can recover the message by aggregating $T$ decryption shares with computation complexity $\mathcal{O}(NK)$. In total, the computation complexity is $\mathcal{O}(NK)$ and the communication complexity is $\mathcal{O}(K)$.

\noindent\textbf{ATASSES}: ATASSES also consists of two rounds. In the first round, each party generates the shares of encryption keys and sends each share to each party with computation complexity $\mathcal{O}(N^2)$ and communication complexity $\mathcal{O}(N)$. In addition, each party also needs to encrypt the message's shares and noises and sends the ciphertexts to the aggregator. The computation complexity is $\mathcal{O}(K)$ and communication complexity is $\mathcal{O}(K)$.
The aggregator needs to compute Lagrange coefficients with communication complexity $\mathcal{O}(N^2)$ and computation complexity $\mathcal{O}(N)$.
In the second round, each party adds $2T$ encryption key shares together and sends the decryption key share to the aggregator, resulting in computation complexity $\mathcal{O}(N^2)$ and communication complexity $\mathcal{O}(N)$. For the aggregator, to recover the decryption key, it needs to compute the linear combination of $T$ decryption key shares with complexity $T\cdot\mathcal{O}(N)=\mathcal{O}(N^2)$. Then it needs to compute the overall ciphertext by summing $T$ ciphertexts up with computation complexity $\mathcal{O}(NK)$ and decrypting it with computation complexity $\mathcal{O}(K)$.
In total, the computation complexity is $\mathcal{O}(N^2+NK)$ and the communication complexity is $\mathcal{O}(N+K)$.

\section{Extension to Fully-Malicious Security}
\label{appendix:fullysecurity}

Below we provide a brief discussion on how to extend our protocol from semi-honest security to malicious security.
Note that the key difference between semi-honest security and malicious security is the adversary's capability. The semi-honest adversary must follow the protocol, while the malicious adversary can operate arbitrarily. Particularly, the actions of malicious adversary can be divided into two categories: one is to abort its corrupted parties and the other is to ask its corrupted parties to mistakenly execute the operations.
Our work mainly discuss how to deal with the first category. Hence, to extend to fully-malicious security, we need to further enable parties to prove the correctness of operations they execute. Thanks to the previous work on maliciously secure ThFHE (e.g., PELTA~\cite{PELTA}) and verifiable secret sharing~\cite{PVSS,PVSS1}, we can adopt their techniques to achieve this goal. Next, we first recap the key techniques in existing works and then discuss how to apply their techniques in our work.

\noindent\textbf{Recap.} Below we recap the PELTA framework and verifiable secret sharing, respectively.

PELTA first categorizes the operations in ThFHE schemes into non-interactive operations and interactive operations. For non-interactive operations such as encryption and homomorphic evaluation, PELTA pointed out that their correctness verification has been addressed by previous works including \cite{Noninteractive1,Noninteractive2,Noninteractive3,Noninteractive4}.
Hence, PELTA put the focus on those interaction operations and identify that these operations share common functionalities.
Namely, the interactive operations in ThFHE schemes can be decomposed into multiple $2$-step interactions. Each interaction consists of the following two steps.

\noindent\textit{Step 1:} In this step, every party $i$ uses its local secret $s_i$ to locally generate a share that can be publicly disclosed. These shares have a common structure. Particularly, a share $b_i$ is computed as a linear equation of the form:
\begin{equation}
    b_i=a\cdot s_i+X+e_i,
\end{equation}
where $a$ is a publicly known value, $s_i$ is the secret value of party $i$, $e_i$ is a newly sampled error term from some distribution, and $X$ is a placeholder that takes different forms in different operations.

\noindent\textit{Step 2:} In this step, the aggregator computes the (weighted) summation of the shares from parties up and outputs a collective value $b$. 

For such $2$-step interactions, they further show how to verify the correctness of these $2$-step interactions via zero-knowledge proof techniques, which can be used for the extension of our work.

Verifiable secret sharing (VSS), introduced by Chor et al.~\cite{PVSS0}, aims to make the (vanilla) secret sharing technique robust against malicious parties. Particularly, it not only prevents a malicious dealer from distributing incorrect shares, but also prevents malicious shareholders from submitting incorrect shares in the reconstruction protocol. VSS will be used when our work relies on secret sharing techniques.

\noindent\textbf{Main Idea.} Recall that the ATASSES construction (see Figure \ref{fig:approxrec}) mainly consists of two parts: one is \textit{ciphertext generation} and the other is \textit{decryption key generation}. The first part involves the BFV encryption (Lines 5--10) executed by each party and the ciphertext summation (Lines 15--16) by the aggregator. We note that these two steps exactly match the above $2$-step interaction and thus can be verified by PELTA's techniques. The second part includes the secret key sharing (Lines 2--4), combination (Lines 12-13), and recovery (Line 14). This part is executed by Shamir secret sharing in our semi-honest version and can be boosted to maliciously-secure construction by verifiable secret sharing such as \cite{PVSS,PVSS1}.

Besides these two parts of ATASSES, there are still several simple operations in ATASSES, including BFV secret key generation (Line 1), Lagrange coefficients computation (Line 11), and BFV decryption (Line 17). The BFV secret key generation is a non-interactive operation and its verification has been discussed in PELTA. As for the Lagrange coefficients computation and BFV decryption, these operations do not involve any secret information and thus every party can check its correctness by simply re-executing these operations. 

In addition, our ThFHE construction has several extra operations beyond the ATASSES. One is secret key sharing (Lines 1--6) in the key generation stage. Similarly, it can be verified by verifiable secret sharing. The other is Phase 1 and Phase 3 in the decryption stage. The Phase 1 asks each party $i$ to output a share $b_i=c_1\cdot \mathsf{skShare}_i+c_0$, which is exactly the Step 1 in the above $2$-step interaction and can be verified by utilizing PELTA's techniques. The Phase 3 is to decode the plaintext $m$ from $b^\prime$. This phase also does not involve any secret information. Hence, it can be verified by asking every party to re-execute this phase.

\end{document}